\documentclass{article}
\usepackage{graphicx} 
\usepackage[export]{adjustbox} 
\usepackage{multirow}
\usepackage{subcaption}
\usepackage{url}
\usepackage{authblk} 
\usepackage[margin=1.2in]{geometry} 

\title{ChatGPT is a Remarkable Tool---For Experts}
\author[1]{Amos Azaria}
\author[2]{Rina Azoulay}
\author[3]{Shulamit Reches}
\affil[1]{School of Computer Science, Ariel University, Israel}
\affil[2]{Dept. of Computer Science, Jerusalem College of Technology, Israel}
\affil[3]{Dept. of Mathematics, Jerusalem College of Technology, Israel}
\date{}

\begin{document}

\maketitle
\begin{abstract}

This paper investigates the capabilities of ChatGPT as an automated assistant in diverse domains, including scientific writing, mathematics, education, programming, and healthcare. We explore the potential of ChatGPT to enhance productivity, streamline problem-solving processes, and improve writing style. Furthermore, we highlight the potential risks associated with excessive reliance on ChatGPT in these fields. These limitations encompass factors like incorrect and fictitious responses, inaccuracies in code, limited logical reasoning abilities, overconfidence, and critical ethical concerns of copyrights and privacy violation.


We outline areas and objectives where ChatGPT proves beneficial, applications where it should be used judiciously, and scenarios where its reliability may be limited.
In light of observed limitations, and given that the tool's fundamental errors may pose a special challenge for non-experts, ChatGPT should be used with a strategic methodology.
By drawing from comprehensive experimental studies, we offer methods and flow charts for effectively using ChatGPT. Our recommendations emphasize iterative interaction with ChatGPT and independent verification of its outputs.
Considering the importance of utilizing ChatGPT judiciously and with expertise, we recommend its usage for experts who are well-versed in the respective domains.
 
%
%


\end{abstract}


\section{Introduction}
\label{section:introduction}

The field of artificial intelligence has rapidly evolved over the years, with natural language processing (NLP) models being one of the most promising areas of research. 
One of the notable developments in this realm is the advent of chatbots and conversational agents \cite{allouch2021conversational}. Owing to their capacity to mimic human responses to text inputs, they have surged in popularity. This rapid rise is greatly attributed to the advancement of large language modules (LLMs), which have significantly enhanced their performance.
LLMs, also referred to as neural language models, are deep learning models that aim to generate human-like text. These models are trained on vast amounts of text data, enabling them to learn patterns, grammar, semantics, and context in a manner similar to human language acquisition.
One such model that stands out is ChatGPT \cite{zhou2023comprehensive}, an AI model with generative capabilities \cite{gozalo2023chatgpt} crafted by OpenAI \cite{brown2020language}. ChatGPT has demonstrated exceptional proficiency across diverse applications, and its latest version, ChatGPT4, exhibits amplified capabilities and is seen as a substantial stride towards achieving artificial general intelligence \cite{bubeck2023sparks}.

The most common ChatGPT uses, as described by the Business Community website\footnote{\url{https://www.business2community.com/statistics/chatgpt}}, are: drafting emails to coworkers, writing a resume or cover letter, summarizing complex topics or long articles, getting answers to questions
without traditional web search, 
writing songs, poetry, and screenplays based on existing content, writing and debugging computer code, translating content into multiple languages, writing essays on any topic, solving math problems, and finding new recipes based on a set of ingredients. In addition, people have also used ChatGPT for some applications that could potentially be risky, such as seeking medical or legal advice. 

ChatGPT has proven to be a valuable tool to promote research. It may serve as a valuable source of inspiration, helping researchers generate ideas,
improve textual expressions, find strategies for conducting research.
By asking ChatGPT questions about relevant analyses or requesting research prompts, researchers can gain insights and guidance for their projects\footnote{\url{https://tilburgsciencehub.com/tutorials/more-tutorials/chatgpt-article/chat-gpt-research/}}.
Furthermore, ChatGPT proves advantageous for various textual operations, such as summarizing, rephrasing, rectifying errors, and translating text, tasks that are of critical importance during the course of any research study.
%

In the realm of education, ChatGPT can be leveraged to create personalized learning experiences for students. 
This is attainable by creating flexible, dynamic content customized to meet each student's individual requirements and by facilitating dialogues and providing responses to their questions.
In creative writing, ChatGPT can assist authors in generating new ideas, enhancing their writing style, and providing valuable ideas. 
In programming, ChatGPT can assist developers in code generation, code debugging, suggesting solution concepts and designs, proposing algorithmic methods, and explaining them. 
In medicine, ChatGPT can be used to analyze medical records, assist in patient diagnosis and treatment, and provide empathetic conversations with the patients.  

In summary, ChatGPT has proven itself as a potent instrument capable of enriching research, boosting efficiency, and refining writing styles across diverse disciplines. By responsibly and transparently employing ChatGPT, we can leverage the full potential of this technology to improve capabilities and promote innovation in different domains.

With ChatGPT's capabilities and limitations in mind, it becomes pertinent to delve into its applications among professional users, which is the primary focus of this paper. It's important to underscore that even in a professional context, deploying ChatGPT can pose significant challenges and risks. Consequently, addressing these challenges and ethical considerations \cite{zhuo2023exploring} is crucial to guarantee safe and effective uses of ChatGPT in the various fields.


Generally, some challenges pertain to the majority of professional uses, while others are specific to particular fields. For instance, in domains where ChatGPT's decisions can have tangible real-world impacts, like patient interactions, transparency should be needed to maintain accountability and prevent inadvertent harm.
In addition, in creative domains and in programming assistance, the issue of copyright related to the source materials that ChatGPT uses for information should also be taken into account.
In the educational field, the fact that a student can produce written work using ChatGPT with little to no effort is a potential issue that educators should consider and address. 
An important concern across various fields is privacy, particularly with regard to the information that ChatGPT learns from its dialogues - an area that is not yet well understood.
Finally, there are concerns about validity and accuracy of the model's predictions and responses. 

In this paper, following a detailed exploration of ChatGPT's capabilities, we address the ethical and practical challenges associated with its use, such as privacy issues, algorithmic biases, and the necessity for transparency in AI technology applications. In doing so, we aim to establish a framework for the responsible incorporation of AI technologies across different sectors, leading to enhanced outcomes for individuals and society collectively.

Several overviews on ChatGPT were already published, discussing its wide range of conversational abilities and common applications \cite{haleem2022era}.
The overview of Zhang et al. \cite{zhang2023one} provides a brief technical description about openAI technology, and in particular, about the technology behind ChatGPT. Similar concept was taken by Ray \cite{ray2023chatgpt}, concentrating on ChatGPT development process, current abilities and achievements, as well as comparing it to other popular LLMs. Both studies provide a list of popular uses in different areas, followed by ChatGPT technical limitations and ethical concerns.

In their extensive review, Ali Khowaja et al. \cite{khowaja2023chatgpt} thoroughly examine a range of concerns related to the use of ChatGPT, including privacy, copyright issues, digital divide, the risk of plagiarism, biased responses,  dissemination of false information, and the lack of accountability.
The authors propose suggestions to counter each of these issues. It's important to note that while Khowaja et al. mainly focus on the ethical challenges at regulatory and institutional level, our research offers practical tools and insights particularly customized for the individual user, with the aim to optimize his/her benefits while adeptly handling the model's practical limitations.


Megahed et al. \cite{megahed2023generative} investigates the effectiveness of ChatGPT in supporting software process improvement practices, learning, and research. Their study finds that ChatGPT performs well for structured tasks like code translation and explaining well-known concepts, but faces challenges with more nuanced tasks such as explaining less familiar terms and generating code from scratch. The researchers suggest that while AI tools like ChatGPT can enhance efficiency and productivity, caution must be exercised as they can produce inaccurate or misleading results. They recommend validating and complementing ChatGPT's use in software process improvement with other methods to ensure accuracy.

A group of 43 experts \cite{dwivedi2023so} in diverse fields published a detailed report concerning the potential of ChatGPT to enhance productivity and offer gains in industries like banking, hospitality, and information technology. In addition, they also discuss limitations, disruptions, privacy concerns, biases, misuse, and misinformation associated with the technology.  several research questions have been proposed, such as 
examining biases in generative AI, determining optimal combinations of human and AI interaction, and addressing ethical and legal issues.

Our paper's unique characteristic lies in its comprehensive portrayal of how ChatGPT can be strategically utilized, within the confines of the system's limitations. We furnish potential users with an array of heuristics and flowcharts for guidance, in addition to proposing strategies for addressing the ethical dilemmas that accompany its usage.
The remainder of this paper is laid out as follows: Section~\ref{section:fields} offers an overview of the possible applications of ChatGPT in a diverse range of fields such as programming assistance, education, mathematics, and healthcare. The technical limitations of ChatGPT are addressed in Section~\ref{section:limitations}. Subsequently, Section~\ref{section:flowchart} introduces techniques and methods that can aid users in maximizing ChatGPT's potential despite its inherent limitations. Finally, we draw conclusions and suggest future directions for research in Section~\ref{section:conclusion}.



\section{Overview of the Potential of ChatGPT Usage in Various Fields}
\label{section:fields}

The advent of advanced chatbots, constructed on the foundations of large language models (LLMs), and significantly fortified by voluminous training data, has ushered in a new era of digital interaction. Their proficiency in understanding and generating natural language has seen them evolve into highly versatile tools, with a broad spectrum of applications spanning numerous industries. Such a tool, exemplified by ChatGPT, possesses tremendous potential that is progressively being recognized, explored, and exploited across an array of sectors in our economy.

In the forthcoming section, we embark on a detailed exploration of the multifaceted utilization of ChatGPT across a range of fields. This includes its potential contributions to scientific research, educational initiatives, programming assistance, mathematical education and problem-solving endeavors, along with its applications in the crucial sector of healthcare.

In each of these respective areas, we delve into an analysis of how ChatGPT can be harnessed to augment human capabilities, thereby leading to more efficient, innovative, and fruitful outcomes. Concurrently, we also discuss the pertinent challenges associated with its use in these fields, highlighting the importance of addressing these challenges to ensure the effective, safe, and ethically sound utilization of this groundbreaking technology. This comprehensive approach aids in providing a holistic understanding of the capabilities, potential applications, and the attendant considerations of using ChatGPT across various professional domains.

\subsection{Research and Academic Usage}
 
ChatGPT serves as a beneficial resource for researcher at all stages of a research project, providing relevant information, guidance, and support to optimize efficiency and effectiveness. During the literature review phase, ChatGPT   can aid researchers by suggesting relevant topics, questions, and methods within their research area and by summarizing important background and related studies \cite{summarizeText,summarizeText2}. This assistance contributes to the construction of a comprehensive literature review and expedites the gathering and analysis of existing literature.

When researchers are in the data collection phase, ChatGPT can share insights on efficient and reliable data collection methods. It can also furnish information on data quality assessment and provide tips to avoid typical data collection errors. When it comes to data analysis, researchers can prompt ChatGPT to propose suitable analysis methods based on the research question and the type of data. It can also provide guidance on interpreting and effectively presenting the results.
ChatGPT proves advantageous even in the final stages of a research project, where it assists in drafting and language editing to enhance readability and grammatical accuracy.

However, it's worth noting that ChatGPT can occasionally make significant errors, which could potentially mislead those who aren't experts in the field. These errors are often coupled with detailed explanations that may initially seem plausible but are ultimately incorrect. ChatGPT version 3.5, for example, has been observed to provide inaccurate quotes, sometimes attributing them to non-existent sources. This issue extends to citations of literary works and Bible verses that don't actually exist. Furthermore, in the realm of proofreading, ChatGPT might commit mistakes when the subject discussed requires specific contextual understanding.

The paper \cite{lund2023chatting}  provides an overview of GPT, focusing on its generative pre-trained transformer model and its versatility in language-based tasks. It explains how ChatGPT utilizes this technology as an advanced chatbot. Additionally, the paper includes an interview with ChatGPT, highlighting its potential impact on academia and libraries. The interview covers various benefits of ChatGPT, including enhancing search, discovery, reference services, cataloging, metadata generation, and content creation. Ethical considerations, such as privacy and bias, are also addressed. Furthermore, the paper explores the feasibility of employing ChatGPT for scholarly paper writing.

The ranking and classification task holds significant importance across various scientific domains, particularly within data science. 
The study of \cite{ji2023exploring} is conducted to assess ChatGPTs ability to rank content. 
To evaluate ChatGPT's ranking capability, a test set with diverse prompts is created. Five models generate responses, and ChatGPT ranks them accordingly. The results demonstrate a certain level of consistency between ChatGPT's rankings and human rankings. This initial experiment suggests that ChatGPT's zero-shot ranking capability could alleviate the need for extensive human annotations in various ranking tasks.
The paper \cite{zhang2022would} deals with  text classification tasks. In particular it considers the task of extracting the standpoint (Favor, Against or Neither) towards a target in given texts. 
Their experiments show that ChatGPT can achieve high performance for commonly used datasets and at the same time, can provide explanation for its own prediction, which is beyond the capability of any existing model. The paper concludes that ChatGPT has the potential to be the best AI model for stance detection tasks in NLP, or at least change the research paradigm of this field.

In an increasingly globalized scientific community, the need for effective and accurate translation is paramount. It aids in the dissemination of research findings across different languages and cultures, promoting a greater exchange of ideas and collaboration among researchers worldwide.
With this in mind, a preliminary study on ChatGPT for machine translation \cite{jiao2023chatgpt} provides insightful observations. The authors examine the performance of ChatGPT across various translation scenarios such as translation prompts, multilingual translation, and translation robustness. Findings indicate that ChatGPT can compete with commercial translation tools like Google Translate, particularly for high-resource European languages.
However, when it comes to low-resource or linguistically distant languages, there's a notable gap in performance. This poses a challenge in scientific contexts where research could be published in any of these less-resourced languages.
Interestingly, the introduction of the ChatGPT4 engine has significantly improved ChatGPT's translation performance. It now stands on a par with commercial translation products, even for those distant languages. This improvement signifies a promising development for the scientific community as it opens up new possibilities for cross-lingual understanding and collaboration.


To summarize, ChatGPT can be a beneficial tool for researchers, assisting in various phases of a research project, from formulating research questions to drafting and language editing. ChatGPT has shown potential in a variety of applications, such as content summarization and ranking, text classification, and machine translation. However, it can also be prone to significant errors, which could mislead non-experts. ChatGPT 3.5 has been observed to provide inaccurate quotes and may struggle with context-specific proofreading tasks.  In addition, while ChatGPT can perform well with certain high-resource languages, it may struggle with low-resource or linguistically distant languages. The introduction of newer versions of ChatGPT have shown marked improvement in these areas.

\subsection{Opportunities and Challenges in Education}

One of the most influential and challenging fields in the world of accelerated technological progress, particularly in the realm of smart chatbots, is education. This dynamic field presents a complex dichotomy; on one hand, it bolsters learners with powerful tools and an enriched reservoir of knowledge, amplifying their intellectual curiosity and acumen.

On the other hand, in this era of unprecedented information availability, the widespread access to a plethora of information, including readily available solutions to exercises, poses a significant challenge to conventional teaching methods. Empowered by the ubiquity of the internet, young learners can easily transcend the confines of the classroom, prompting educators to reassess and fine-tune their pedagogical strategies. This juncture of opportunities and predicaments in the technology-enhanced educational landscape triggers a robust conversation on the future trajectory of teaching and learning.

Two additional considerations need to be addressed. The first pertains to the educational challenges posed by potentially partial, unreliable, or incorrect responses students might receive when interacting with chatbots. However, this concern is expected to diminish as technology progresses. The second consideration revolves around the changing role of educators in a world increasingly dominated by AI, particularly conversational technology such as ChatGPT. This concern grows more acute as technological advancements continue. The abilities of ChatGPT, which include providing detailed explanations across a wide range of fields, have led to conjecture about its potential to supplant traditional teaching methods in some areas.

In  \cite{opara2023chatgpt}, the authors conduct a literature assessment on the educational implications of artificial intelligence, focusing on OpenAI's ChatGPT. The study highlights the advantages of ChatGPT, an AI-powered tool, in improving student access to information by providing immediate, tutor-like responses, particularly benefiting self-paced learners. However, it also acknowledges its limitations, such as occasionally producing incorrect or illogical responses, and sensitivity to input phrasing. For example, while it might not answer correctly with one phrasing, a slight rephrase can elicit a correct response. 

In \cite{lo2023impact}, an extensive literature review is carried out within the initial three months following the release of ChatGPT. The review rigorously examines the capabilities of ChatGPT, its potential roles in the educational sector, and pertinent concerns associated with its use. The analysis underscores the potential of ChatGPT as a valuable tool for instructors, providing a basis for the development of course syllabi, teaching materials, and assessment tasks. However, the study also addresses concerns regarding the accuracy and reliability of the content generated by ChatGPT, as well as the challenge of plagiarism prevention. The study emphasizes the need for prompt action by educational institutions to update their guidelines and policies concerning academic integrity and plagiarism prevention. Additionally, it advocates for the training of instructors on the effective use of ChatGPT while equipping them with the skills to identify instances of student plagiarism.

AI-based comprehension abilities could serve as valuable educational tools for detecting and analyzing classroom social dynamics and students' emotional states. Phillips et al.'s study \cite{phillips2022exploring} demonstrates the capabilities of ChatGPT-3 in summarizing student interactions in a game-based learning setting and its precision in identifying various emotions and behaviors. Despite limitations such as understanding context and detecting hyperbole, GPT-3 can provide insightful interpretations of collaborative dynamics, potentially reinterpreting seemingly unproductive conversations as crucial for managing negative emotions. Teachers can employ ChatGPT-3 to monitor real-time student discussions, address students experiencing difficulties, and observe progress. The study also suggests that teachers act as intermediaries between the AI's interpretations and educational decisions, thus mitigating the risk of model failure.

The study by \cite{farrokhnia2023swot} conducted a SWOT (Strengths, Weaknesses, Opportunities, and Threats) analysis of ChatGPT in the education sector. 
The strengths of ChatGPT identified in the study include its ability to provide plausible and credible responses compared to other AI tools, its self-improving capability, personalized responses, and its aptitude for understanding complex inquiries and delivering real-time relevant answers. The opportunities presented by ChatGPT in education revolve around increased accessibility to information. It can efficiently find and summarize relevant information, making it easier for students to access detailed knowledge quickly. Moreover, ChatGPT has the potential to offer personalized support and feedback to students at varying levels of complexity. It can also stimulate critical thinking by challenging students with tailored sets of questions corresponding to their proficiency level.

However, the study also highlights several weaknesses of ChatGPT. One weakness is its lack of deep understanding. While it recognizes patterns and generates plausible responses, it does not possess a comprehensive grasp of the underlying concepts. ChatGPT also struggles with evaluating the quality of responses. Additionally, there are inherent risks of biases and discrimination. Biases in training data, algorithmic design, and societal context can perpetuate biases and discrimination within ChatGPT's outputs. Moreover, the use of ChatGPT raises concerns about academic integrity, as it can be exploited for cheating in online exams and compromise assessment security. Ethical issues such as the provision of fake information and similarities to existing sources are also potential challenges associated with ChatGPT. Lastly, the use of ChatGPT by students may lead to a decline in higher-order cognitive skills, including creativity, critical thinking, reasoning, and problem-solving.

Delving into each domain separately brings with it a unique blend of opportunities and challenges. We proceed by dissect research that explores the outcomes of incorporating ChatGPT into an array of educational sectors. This nuanced approach aids in illuminating the specific implications of AI integration within these distinct fields.
 
Cotton et al. \cite{cotton2023chatting}  examines the opportunities and challenges of using ChatGPT in higher education, and discusses the potential risks and rewards of these tools. The paper also considers the difficulties of detecting and preventing academic dishonesty, and suggests strategies that universities can adopt to ensure ethical and responsible use of these tools. According to this paper,  the integration of chatAPIs and GPT-3 in higher education holds significant potential for enhancing student engagement, collaboration, and accessibility. The use of chatAPIs enables asynchronous communication, timely feedback, group work, and remote learning support. Similarly, GPT-3 offers valuable applications such as language translation, summarization, question answering, text generation, and personalized assessments. However, the adoption of these tools also presents challenges and concerns, particularly regarding academic integrity and plagiarism. The use of chatAPIs and GPT-3 can potentially facilitate cheating, and distinguishing between human and machine-generated writing can be challenging. To ensure ethical and responsible usage, universities need to carefully evaluate the risks and rewards associated with these tools. This involves developing policies and procedures, providing training and support for students and faculty, and implementing robust methods to detect and prevent academic dishonesty. By addressing these challenges, universities can harness the opportunities offered by chatAPIs and GPT-3 while upholding the integrity of their assessments and maintaining the quality of their educational programs.

Another potential usage of ChatGPT is to guide students and teach them some mathematical issues. In \cite{pardos2023learning} the authors check the learning gain evaluation of ChatGPT by comparing the efficacy of its hints with hints authored by human tutors, across two algebra topic areas, Elementary Algebra and
Intermediate Algebra. All their experiments produced learning gains; however, they were only statistically significant among the manual hint conditions.
Manual hints produced higher learning gains than ChatGPT hints
in both lessons and these differences were statistically significantly
separable. They found out that the technology,  in its current form, still requires human supervision. Their result showed 30 $\%$ rejection rate of produced hints based on quality, suggesting that All of the rejected hints were due to containing the wrong answer or wrong solution steps. None of the hints contained
inappropriate language, poor spelling, or grammatical errors.

The research conducted in \cite{qadir2022engineering} explores the use of generative AI and ChatGPT techniques in the context of engineering education. The authors engage ChatGPT by posing various questions related to this topic and analyze its responses. They ultimately conclude that ChatGPT and similar AI language models hold significant potential as convenient and beneficial tools for both students and teachers in engineering education. These models have the ability to generate text resembling human-like conversation, provide answers to questions, compose essays, and assist with homework assignments. Potential applications encompass language editing, virtual tutoring, language practice, generating and solving technical and non-technical queries, as well as aiding in research tasks. However, it is crucial to recognize that ChatGPT and other AI language models are not flawless and can produce errors or furnish incorrect information. Therefore, it is imperative to exercise caution when using these tools and establish community guidelines and standards to ensure their fair and responsible use. 

The paper \cite{cribben2023benefits} discuss the benefits
and limitations of ChatGPT in Business education and research specifically focusing on management science, operations management, and data analytics. The study considers how professors and students can utilize ChatGPT in these areas. Professors can leverage ChatGPT to design courses, develop syllabi and content, assist with grading, and enhance student comprehension. On the other hand, students can rely on ChatGPT to explain intricate concepts, aid in code creation and debugging, and generate sample exam questions. The primary strength identified in this analysis is ChatGPT's proficiency in writing and debugging code, making it particularly valuable for educational and research purposes. However, it is essential to acknowledge that ChatGPT does have limitations, including occasional errors and a requirement for a deeper or advanced domain knowledge. Additionally, the discussion surrounding ChatGPT in business education and research raises concerns regarding potential biases and issues related to plagiarism.

We proceed by describing two additional studies that propose various strategies to address the integrity challenges.
The study conducted by Ryznar et al. \cite{ryznar2023exams}, explores diverse methods to maintain the integrity of examinations in the era of open AI technologies, including ChatGPT. This research presents a comprehensive range of strategies to safeguard exam integrity, which encompass high-tech solutions such as video proctoring and specialized exam software, as well as low-tech measures like time constraints and meticulous course design. By presenting this array of strategies, the study aims to illuminate effective practices for exam administration that uphold integrity despite the widespread use of technologies like ChatGPT.

In a separate study, Shidiq et al. \cite{shidiq2023use}, scrutinize the impact of the ChatGPT system on students' writing skills, with a particular focus on creativity. While the capacity of ChatGPT to generate responses based on inputted keywords has potential benefits for education and learning, the researchers note that not all aspects of this technology necessarily contribute effectively to the development of a diverse range of student skills, including creative writing. To address this, the paper underscores the importance of educators implementing strategies that go beyond online learning tools, which students may exploit while completing assignments. One proposed approach involves using paper as a platform for task development, serving as a mechanism for process control and specific assessment of creative writing tasks. When this method is implemented, it enables teachers to offer a structured framework that can guide students and assess their progression in creative writing.

In conclusion, ChatGPT proves to be a valuable tool for lecturers and instructors to structure their lessons, offering personalized feedback and serving as a starting point for course development. It surpasses mere Google search responses, providing improved student access to information and delivering personalized support with a personal touch. Moreover, it enhances student engagement, collaboration, and accessibility, enabling students to rely on it for explanations, code generation, and sample test questions. Overall, ChatGPT empowers instructors and students across various educational domains.

However, there are certain limitations and concerns associated with ChatGPT. One concern is its potential to generate incorrect or fabricated information, which can have implications for academic integrity. Despite its ability to produce plausible responses, ChatGPT lacks a deep understanding and may provide responses that are incorrect or illogical. It is also sensitive to changes in input phrasing or multiple attempts at the same question, as a slight rephrasing can lead to an accurate response. Additionally, not all aspects of ChatGPT effectively contribute to the development of various student skills, particularly in the realm of creative writing. While it can recognize patterns and generate plausible responses, it lacks a comprehensive grasp of underlying concepts and struggles with evaluating the quality of its responses.

Moreover, there are inherent risks of biases and discrimination in ChatGPT's outputs, stemming from biases in training data, algorithmic design, and societal context. Furthermore, the use of ChatGPT raises concerns about academic integrity, as it can be exploited for cheating in online exams and compromise the security of assessments. Ethical issues, such as the provision of fake information and similarities to existing sources, pose additional challenges associated with ChatGPT.

\subsection{Programming Assistance}
ChatGPT can serve as a valuable tool for programmers throughout the software development process, providing suggestions, guidance, and feedback that contribute to enhanced efficiency and effectiveness in programming. Numerous online resources offer lists of practical applications for ChatGPT, along with accessible short courses for learning its functionalities. Numerous articles and blogs also delve into specific use cases and provide detailed examples \cite{KenJee3, makeuseof2023, levelup2023, typefully2023, tian2023chatgpt, surameery2023use, help-developers}.

The key principle in utilizing ChatGPT for programming assistance involves supplying it with appropriate prompts to achieve desired outcomes \cite{LLMPrompting, real-programming, coding-prompts-greataiprompts}. However, it is important to note that, as demonstrated in previous publications and in this study, effective use of ChatGPT for professional development in software engineering requires significant proficiency in both interacting with AI tools such as ChatGPT and in software development skills.

Toam et al. \cite{tian2023chatgpt}
conducted an empirical study on ChatGPT to assess its potential as a programming assistant. They focused on three code-related tasks: code generation, program repair, and code summarization. For code generation, ChatGPT performed well on common programming problems but struggled with generalization to new problems. In program repair, ChatGPT achieved competitive results compared to a state-of-the-art tool. However, it had a limited attention span and performed worse when provided with unrelated prompts. The study highlights the importance of prompt engineering and provides insights into ChatGPT's practical applications in software engineering.

Xu Hao, the Head of Technology for China at Thoughtworks, demonstrated the use of ChatGPT in the programming process \cite{LLMPrompting, real-programming}. He began by setting the context for the application and defining the desired code structure through a prompt, then detailed the system's common implementation strategy and requirements. Xu used ChatGPT to create a solution outline without producing code, which he reviewed and revised to align with the architectural vision. Next, he had ChatGPT generate a detailed plan, including component names, methods, and props. He also requested test examples and code implementation from ChatGPT, prioritizing tests first. Lastly, he evaluated the generated code, refining it using his expertise and the provided examples. In essence, Xu showcased how ChatGPT, when used effectively in conjunction with expert knowledge, can significantly enhance the software development process.

The primary strategy recommended by experts for generating innovative results involves a method called ''Knowledge Generation". This approach entails formulating lengthy and complex prompts to elicit functional code \cite{LLMPrompting}. These prompts encompass architectural descriptions, design guidelines, and step-by-step action plans, emphasizing the importance of testing and thorough code inspection. Experts highlight that programming with ChatGPT is a unique form of coding that relies heavily on a comprehensive understanding of architecture, system knowledge, and rigorous testing. While this approach offers potential time-saving benefits and automation of certain tasks, it still requires learning and mastery as a skill.
The blog \cite{coding-prompts-greataiprompts} provides a comprehensive collection of commonly used prompts to effectively utilize ChatGPT for programming assistance. Additionally, \cite{improving-skills-10} offers ten insightful suggestions for intelligent and efficient utilization of ChatGPT.

The rest of this section highlights various practical applications of ChatGPT for programming objectives. These applications draw inspiration from various authors, bloggers, and our own experiences.

{\bf Enhance programming skills:} \cite{lab-tutorial,writing-code-trustworthy}: 
ChatGPT offers a time-saving solution for learning programming languages by aggregating information from various sources and presenting it in a single interface. It offers code explanations, alternative approaches, and serves as a real-time tutor. Programmers may use ChatGPT as an online teacher to enhance their programming skills \cite{levelup2023}. This can be done through code explanations and by introducing relevant technologies, coding methods, and software packages. Additionally, ChatGPT can provide feedback and recommendations on the code, aiding in understanding errors and potential enhancements \cite{lab-tutorial}.

Moreover, ChatGPT's ability to generate code and provide valuable explanations enables developers who are unfamiliar with a programming language or framework to quickly catch up without spending excessive time on the fundamentals. This is particularly valuable for beginner programmers seeking to accelerate their learning process \cite{help-developers}.

{\bf Information gathering:}
ChatGPT is able to provide relevant information and explanations on complex programming concepts \cite{typefully2023}. By utilizing ChatGPT, developers can quickly obtain answers to specific technical questions, access relevant code samples, and save time searching for solutions or examples on the internet. In particular, ChatGPT can be used to explore libraries and resources, especially in the context of intelligent web page data extraction \cite{writing-code-zdnet}.

{\bf Code explanation:}
When asked to explain code, ChatGPT has the capability to provide detailed natural language explanations \cite{typefully2023}. However, it's important to note that, based on our experience, these explanations can sometimes be technical and may not always capture the intended meaning of the code. For example, if the code contains inappropriate variable names, the explanations may focus on those names instead. Additionally, there is a possibility of incorrect explanations, as shown in Section~\ref{sec:incorrect}, so caution should be exercised when relying on the chatbot for code understanding \cite{maximize-coding-potential}.

{\bf Code generation:}
ChatGPT is a powerful tool for developers, leveraging natural language processing to understand and interpret developer needs. It can save programmers time by assisting with repetitive tasks and boilerplate code \cite{help-developers}. It is useful for scaffolding \cite{typefully2023} and generating foundational code elements, helping overcome the Cold Start Problem \cite{KenJee3}. 

ChatGPT's programming capability extends to various programming languages, both old and new \cite{writing-code-zdnet}, and it can even convert code from one language to another \cite{maximize-coding-potential}. This enhances its coding potential and enables seamless transitions between languages. In addition, ChatGPT proves adept at generating website code when provided with detailed prompts \cite{levelup2023}, making it a valuable tool for web development tasks. It also demonstrates understanding of Linux \cite{levelup2023,maximize-coding-potential} and SQL \cite{maximize-coding-potential} command lines, allowing it to interpret and respond to queries and commands in these domains. This expands its usefulness in assisting with Linux-based operations and interacting with SQL databases.


While ChatGPT has been rapidly adopted in the industry, caution is advised when using AI-generated code in critical software systems \cite{makeuseof2023}. It is recommended to use ChatGPT as a companion tool, with human programmers retaining control over the development process,
since trying to outsource the entire software process to Chatbot can be a recipe for disaster \cite{makeuseof2023}. 

Although ChatGPT's coding level is observed to be comparable to that of a first-year programming student, with a lack of expertise and diligence \cite{writing-code-zdnet}, it still provides value by assisting with code writing and information lookup. While it may not be able to handle complex projects on its own, it can enhance productivity and efficiency in software development \cite{maximize-coding-potential}.

"Conversational Coding" refers to the process of using ChatGPT for code generation, allowing developers to leverage their critical thinking skills without the need to directly translate their thoughts into code \cite{better-coding-towardsdatascience}. By prompting ChatGPT with their intentions and engaging in dialogue, the developers can collaborate with the model to refine and improve the generated code. If any issues arise, the developers can report them to the chatbot, prompting it to provide updated code. This iterative process enables effective collaboration between developers and ChatGPT to achieve desired results.


A suggested approach for utilizing ChatGPT in data science \cite{KenJee3} is to use it as a high-order functional unit. ChatGPT can generate code based on specified parameters and adapt it for different learning methods, enabling code reuse, automation, and time-saving in adapting code for various models. This usage of ChatGPT resembles the behavior of a function, providing functionality tailored to specific requirements.


In Section~\ref{section:flowchart}, we present some flowcharts that can be useful for problem-solving with ChatGPT. These flowcharts can be used to describe iterated sessions of requests, code generation, careful review, and code refinements, leading to the desired results.

{\bf Code debugging:} 
One of the common uses of ChatGPT, which is widely suggested by programming blogs, is programming debugging \cite{typefully2023,levelup2023,help-developers,surameery2023use}.
They note that ChatGPT is a powerful tool for identifying coding errors, ranging from simple syntax mistakes to complex logic errors.
 Developers can provide problematic code to obtain error detection assistance and further guidance by describing desired outcomes and current outputs \cite{makeuseof2023}.

ChatGPT's human-like interface provides succinct explanations and integrates additional hints, resulting in a significantly improved success rate in resolving programming issues \cite{maximize-coding-potential}. Detailed information, such as programming language and environment, enhances bug hunting with ChatGPT \cite{makeuseof2023}. It examines bugged code, suggests actions for bug identification and correction, and proposes modifications to enhance readability and maintainability, reducing the occurrence of bugs and expediting development cycles \cite{help-developers}. 
However, since mistakes are still faced, careful use of ChatGPT is important, and its outputs should be validated \cite{surameery2023use}.

In an experimental evaluation by Sobania et al. \cite{sobania2023analysis}, ChatGPT's bug fixing abilities were compared with standard methods. ChatGPT performed competitively with deep learning-based approaches and outperformed traditional program repair methods, achieving a 77.5\% success rate. The study highlighted the value of human input in improving an automated program repair system, with ChatGPT facilitating collaboration. The authors acknowledged that the mental effort required to verify ChatGPT's answers can be significant, suggesting the integration of automated approaches to provide hints and verify responses, thereby enhancing ChatGPT's performance and making it a practical tool for software developers in their daily tasks.

Surameery et al. \cite{surameery2023use} discuss the characteristics of ChatGPT in providing debugging assistance, bug prediction, and bug explanation. They highlight its potential in these areas while acknowledging the importance of using other debugging tools and techniques for validation. The paper concludes by suggesting that ChatGPT can be a valuable component of a comprehensive debugging toolkit, complementing other tools to effectively identify and fix bugs.

To summarize, ChatGPT is found to be a powerful tool for programming debugging, capable of identifying coding errors and providing guidance from syntax mistakes to complex logic issues. The human input is valuable in verifying ChatGPT's answers and providing additional context or insights that may aid in bug resolution. By combining the automated assistance of ChatGPT with human expertise and validation, developers can effectively collaborate to identify and fix bugs in a more efficient and accurate manner.

{\bf Code optimization:} 
ChatGPT possesses the ability to analyze user-provided code and suggest improvements in terms of efficiency, security, and readability \cite{help-developers}. Developers can prompt ChatGPT to propose optimization techniques or generate optimized versions of code, with the AI model providing explanations for its corrections and highlighting areas of potential improvement \cite{makeuseof2023}.

Moreover, ChatGPT can generate alternative code that enhances efficiency, scalability, and performance across different programming languages and patterns, leading to more effective and maintainable code \cite{help-developers}. It can also rewrite code to improve coding style, simplicity, and other desired aspects based on the specific requirements and preferences of programmers \cite{typefully2023}.

For optimal results, it is advisable to formulate precise and specific queries when interacting with ChatGPT, as the suggested corrections and recommendations depend on both the inputted code and the context of the query \cite{levelup2023}.


{\bf Data formatting and data creation:}
ChatGPT can also be utilized for data formatting tasks, such as structuring data into specific formats like CSV or JavaScript objects, and generating filler content \cite{makeuseof2023}. It has the capability to create regular expressions \cite{makeuseof2023} and generate formatted content in various formats like LaTeX, HTML, and others. Furthermore, ChatGPT can generate random numbers following specific statistical distributions, which can be beneficial for data augmentation in training machine learning models \cite{dai2023chataug}.

{\bf Test cases generation:}
One effective approach to ensure bug-free and robust code that handles exceptions and edge cases is to write unit tests. ChatGPT can be a useful tool for this task as well \cite{makeuseof2023}. Developers can leverage ChatGPT to assist them in writing test cases for specific functions by providing the relevant code and detailed instructions \cite{typefully2023}. ChatGPT can generate test inputs and expected outcomes, covering various code paths and edge cases. It can also aid in creating concise and comprehensible documentation for test cases, including inputs, predicted outcomes, and pass/fail conditions \cite{help-developers}. While ChatGPT can automate the process of writing test cases, it is still advisable to review and verify the generated test cases rather than relying solely on them.

{\bf Project Documentation:}
ChatGPT is a versatile tool that excels at generating comprehensive code documentation \cite{typefully2023}. It can provide detailed information, incorporate usage examples, and explain code in plain English to assist developers in understanding its functionality and purpose  \cite{levelup2023}. Leveraging its natural language processing abilities, ChatGPT accurately identifies code requirements and generates informative documentation, including automatic comments, to aid future development \cite{help-developers}.


{\bf Code translation:}
ChatGPT offers valuable assistance in code translation, enabling the smooth transfer of code from one programming language to another \cite{typefully2023,levelup2023}. This functionality proves useful when encountering programming challenges in one language and seeking solutions in different languages \cite{makeuseof2023}. Specifically, it provides significant benefits to users who need to migrate applications from a mainframe to a PC-based platform or when dealing with unsupported languages \cite{posey2023}.

{\bf A general comment}
As mentioned by \cite{writing-code-trustworthy} and other programming experts, ChatGPT is not infallible. Like any AI system, it can make mistakes and may exhibit unwarranted confidence in those mistakes. Therefore, it is recommended to remain vigilant and continue to verify, test, and debug its output.

Another limitation of ChatGPT is its lack of comprehensive context \cite{writing-code-trustworthy, trust-bbc}. While it can provide code snippets or even entire files, it lacks an understanding of specific conventions, best practices, or project requirements unique to your company or project. It cannot anticipate how the code will interact with other components or consider critical aspects such as performance, security, privacy, and accessibility. Hence, the ultimate responsibility for the code lies with human developers.

However, as stated by \cite{levelup2023}, ChatGPT cannot entirely replace programmers as programming requires various skills such as problem understanding, solution design, testing, domain knowledge, and communication.

By effectively utilizing ChatGPT as a valuable tool, programmers can focus on critical thinking and human creativity \cite{levelup2023}. It is advisable to master ChatGPT as a companion to enhance programming work and increase competency, rather than relying solely on it for coding tasks. While ChatGPT provides a powerful means to expedite the coding process, it should not be seen as a magical tool that eliminates the need for human effort and understanding \cite{makeuseof2023}.

To summarize, ChatGPT offers valuable applications for programmers across the software development process. It can enhance programming skills, provide information, explain code, assist in code starting and generation, aid in code debugging and optimization, format and create data, generate test cases, document projects, and translate code from one programming language to another. It is important to utilize ChatGPT effectively, considering prompt engineering and expert knowledge, and to validate its outputs. While ChatGPT has limitations and cannot replace human programmers, it serves as a powerful tool to expedite coding tasks and enhance productivity when used as a companion in the programming workflow.

\subsection{Mathematics Tasks}


Mathematical materials are widely available on the internet, and unlike fields such as current affairs that are constantly evolving, mathematics is based on fundamental principles that do not change on a day-to-day basis. Consequently, it would be reasonable to expect that ChatGPT, that has been trained on a vast corpus of text data, which includes a significant amount of mathematical concepts and operations, would be able to comprehend and analyze mathematical problems at least at the undergraduate level, such as linear algebra and calculus. However, in practice, it has been found that ChatGPT's comprehension levels are far from satisfactory, as we will explain in detail below.

In fact, ChatGPT has the ability to perform various mathematical tasks, including solving standard equations, simplifying expressions, computing derivatives and integrals, and performing basic arithmetic operations.
ChatGPT can also generate mathematical expressions and equations, and is capable of answering questions related to mathematical concepts and theories. Additionally, ChatGPT can provide step-by-step explanations and examples to help users better understand mathematical concepts and problem-solving strategies.


However, according to our study and the related work, ChatGPT demonstrates overconfidence in the mathematical field that surpasses its actual capabilities. Its logical analysis is lacking, and it struggles to comprehend algebraic representations that include parameters or descriptions of group members both algebraically and verbally. When answering mathematical questions, ChatGPT provides detailed responses, but with incorrect or insufficient reasoning. Furthermore, it often displays incorrect answers with confidence, particularly in complex arithmetic calculations. Azaria \cite{azaria2022chatgpt} demonstrates the limitations of ChatGPT in processing complex mathematical expressions and its tendency to produce random digits are explored. The paper highlights several difficulties ChatGPT faces in tasks such as multiplying large numbers, computing roots and powers of numbers (especially fractions), and adding (or subtracting) irrational numbers like $\Pi$ or $e$. The study notes that ChatGPT is unaware of its limitations and may simply generate random digits when faced with a complex mathematical expression. To analyze ChatGPT's frequency of digit output, the researchers subjected it to mathematical queries that resulted in irrational numbers. The paper also includes an appendix that discusses ChatGPT's responses to common social experiments, demonstrating its tendency to answer like humans.

The paper \cite{frieder2023mathematical} investigates the mathematical capabilities of ChatGPT.
It tested whether ChatGPT can be a useful assistant to professional mathematicians by emulating various use cases that come up in their daily activities.
When they 
evaluate its performance against other mathematical datasets, they found out that that 
ChatGPT's mathematical abilities are significantly below those of an average mathematics graduate student and that ChatGPT often understands the question but fails to provide correct solutions.
They conclude that ChatGPT is not yet ready to deliver high-quality proofs or calculations consistently.
To summarize, the current version of ChatGPT has some limited ability to quote definitions, theorems, and proofs, and it can solve and explain some types of known mathematical equations and challenges, but it fails when some deeper logical understanding is required. 

To summarize, ChatGPT is adept at solving basic mathematical problems and explaining concepts, but it struggles with complex operations and often provides incorrect answers confidently. Its logical reasoning abilities fall short, particularly in algebraic representations and comprehension. As a potential solution, future versions of ChatGPT could be trained more intensively on complex mathematical datasets, with a focus on improving logical reasoning abilities and reducing overconfidence in problem-solving. Additionally, implementing mechanisms for it to be aware of and communicate its limitations could increase its utility and user trust.

\subsection{Healthcare}


ChatGPT has distinguished itself uniquely in the field of medicine and healthcare, exhibiting substantial potential in assisting medical professionals due to its advanced capabilities. As demonstrated in a study by Kung et al. \cite{kung2023performance}, ChatGPT's aptitude in processing intricate medical and clinical information resulted in a commendable performance on the United States Medical Licensing Examination (USMLE), often surpassing the passing threshold. Intriguingly, despite its general content training, ChatGPT outperformed PubMedGPT, a model specifically trained on biomedical literature, indicating that a combination of broad-based and domain-specific models could substantially boost accuracy. 
In the study of Johnson et al., \cite{johnson2023assessing}
33 physicians across 17 specialties generated 284 medical questions and graded ChatGPT-generated answers to these questions for accuracy. They found out that ChatGPT generated accurate answers and got completeness scores across various specialties, question types, and difficulty levels. 

Nonetheless, despite its impressive proficiency in the medical field, it's crucial to understand the potential limitations of its application, especially within such a high-risk domain where issues of responsibility are heightened. Of particular concern is the use of ChatGPT as a tool for medical consultation among non-professional users. It's vital to underline that this AI should not be regarded as a substitute for professional medical advice.

Sallam \cite{sallam2023chatgpt} provides an exhaustive review of the potential applications and drawbacks of ChatGPT within healthcare education, research, and practice. On the beneficial side, Sallam points out ChatGPT's contribution to enhancing scientific writing, promoting research versatility, facilitating data analysis, aiding in literature reviews and drug discovery, streamlining workflow, and fostering personalized learning in healthcare education. 

However, the review also underscores various potential pitfalls including ethical and legal dilemmas, copyright and transparency issues, risk of bias and plagiarism, possible generation of inaccurate content, cybersecurity threats, and the danger of information overload, or 'infodemics'. 

Sallam further stresses concerns about the use of ChatGPT in healthcare, specifically its lack of personal and emotional perspectives crucial for healthcare delivery and research. The review also points out potential challenges in healthcare education, such as the quality of training datasets possibly leading to biased and outdated content. Lastly, limitations of ChatGPT, such as its current inability to handle images, underperformance in certain areas (as illustrated by its failure to pass a parasitology exam for Korean medical students), and potential plagiarism issues, are discussed.

Biswas \cite{biswas2023role,biswas2023chatgpt} examined the uses of ChatGPT both in public health and in medical research. In \cite{biswas2023role}, the potential applications of ChatGPT in public health are scrutinized. The paper underscores ChatGPT's capabilities, such as dispensing information on public health issues, fielding questions on health promotion and disease prevention strategies, clarifying the role of community health workers and health educators, debating the impact of social and environmental factors on community health, and providing insights about community health programs and services. Nevertheless, using ChatGPT in this area isn't without constraints, including issues of limited accuracy, inherent biases and data limitations, context insensitivity, diminished engagement, and the lack of direct communication with health professionals.

In another study, Biswas \cite{biswas2023chatgpt} explores the potential advantages and ethical concerns associated with using ChatGPT in medical research. Among the ethical issues raised are questions of authorship, accountability, and authenticity. Legal challenges, particularly around copyright infringement and specific field regulations, are also noted. Potential drawbacks, such as stifled innovation due to repetitive text generation and decreased student engagement, are considered. Moreover, concerns regarding accuracy and bias in AI-generated text are brought to light, as AI models trained on large, potentially biased datasets could inadvertently perpetuate or amplify such biases.


Despite ChatGPT's ability to provide medical information, respond to medical inquiries, and suggest differential diagnoses for common symptoms, substantial concerns persist about its decision-making process and outdated information. As highlighted by Arif et al. \cite{arif2023future}, employing ChatGPT in healthcare presents significant obstacles, predominantly due to its decision-making process and outdated data. The authors question ChatGPT's lack of critical thinking and its habit of presenting information redundantly and irrationally. The model's training data, only updated until 2021, and its restricted access to major medical databases such as PubMed and Cochrane, are noted as significant hurdles. These constraints not only limit its applications to tasks like abstract writing, but they also raise questions about its overall credibility.

In summary, ChatGPT has exhibited considerable capabilities in the realm of healthcare, offering considerable assistance to professionals by deciphering intricate medical information. It has displayed an admirable performance in medical examinations and has proven accurate in addressing various medical queries across numerous specialties. 

However, there are ongoing concerns about its lack of personal and emotional insights crucial to healthcare delivery, potential biases in training datasets, difficulties with image handling, and possible plagiarism. In medical research, it  grapples with ethical and legal challenges, such as issues concerning authorship, authenticity, and copyright infringement. 

Additional concerns encompass the possibility of generating inaccurate content and drawing incorrect conclusions. Moreover, the reality that the training data doesn't extend beyond some constate date serves to restrict its practical usage and casts doubt on its dependability.

As a result, while ChatGPT may offer valuable perspectives in public health and medical research, it's essential to understand that it should not replace professional medical advice due to concerns surrounding its decision-making capabilities, potential bias, and outdated information.

\section{Technical Limitations and Ethical Concerns}
\label{section:limitations}

As established in previous research \cite{bang2023multitask, taecharungroj2023can, kocon2023chatgpt}, ChatGPT has been found to have several limitations. These limitations include the potential to provide incorrect responses, generate inaccurate code, rely on outdated information, have limited logical reasoning capabilities, lack self-correction abilities, and display overconfidence. Additionally, there is a concern about the tendency of ChatGPT to produce biased or inappropriate responses. 


Given ChatGPT's ability to provide in-depth explanations, supported by examples and references, it becomes challenging 
to navigate its occasional inaccuracies, critical errors, and fundamental mistakes.
 Therefore, ChatGPT should be utilized with caution, and its outputs should be independently verified using other reliable tools. These limitations position it as a tool that is especially beneficial for those with expertise in the respective fields.

In this section, we will discuss some of the inherent limitations of ChatGPT, and in Section \ref{section:flowchart}, we will present methodologies and strategies that can help address these limitations responsibly and with caution.
%









\subsection{Incorrect Responses - Confidently}
ChatGPT faces a noteworthy challenge related to its occasional delivery of inaccurate information while projecting an unwavering sense of certainty. This limitation significantly hampers the platform's effectiveness, necessitating users to approach its responses with caution and independently verify the information provided. This becomes especially crucial in domains where there are clear-cut correct and incorrect answers. For instance, in medical or scientific contexts, the provision of incorrect information by ChatGPT, coupled with its high-level language explanations and confident demeanor, can lead to confusion among students and non-experts. 

\subsubsection{Incorrect Responses}
\label{sec:incorrect}

Previous studies have revealed that ChatGPT is susceptible to cognitive biases commonly observed in humans \cite{azaria2023toohuman}. Additionally, it has been observed that ChatGPT can generate hallucinated content \cite{alkaissi2023artificial}. This issue arives consistently when  ChatGPT3.5 attempts to provide links and references.

















\subsubsection{Inaccurate Code}
\begin{figure} 
    \centering
    \includegraphics[width = 12cm]{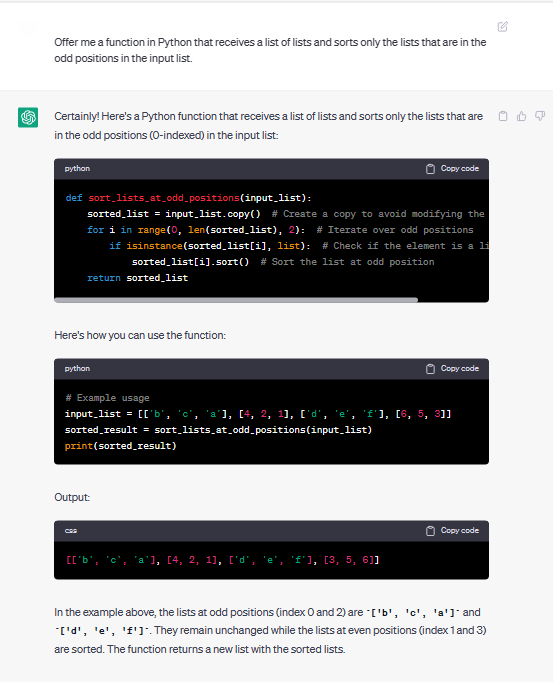} 
    \caption{Incorrect code provided by ChatGPT for a simple sorting task.}
    \label{figure:sorting} 
\end{figure}
ChatGPT often provides inaccurate code. Furthermore, ChatGPT often provides different code snippets for similar and even identical prompts. Therefore, it may provide correct code for a query once, but provide incorrect code when asked a second time (See example in Figure \ref{figure:sorting}). 
Often, ChatGPT attempts to explain the code that it has generated. However, this explanation is sometimes accurate, but sometimes incorrect.
Thus, when utilizing ChatGPT for code generation, it is crucial to pay attention to the quality of the generated code. As we later discuss, it is possible to rephrase a query and ask ChatGPT to solve the same problem multiple times. One can then study the differences between the different responses and attempt to determine whether the responses can be trusted. 
In order to ChatGPT code trustworthy, we ask ChatGPT the following query: Please provide 30 variations for the following text: ``Offer me a function in Python that receives a list of lists and sorts only the lists that are in the odd positions in the input list.'' All the query variations generated by ChatGPT were correct. 
Subsequently, we utilized each of the 30 distinct query variations generated by ChatGPT as prompts in separate chat windows. We then proceeded to evaluate the resulting function code, the accompanying main program that invokes the function, the output examples, and the textual explanations provided by ChatGPT.
For the above experiment, run on ChatGPT3.5, 93.3\% of the produced functions were correctly produced, as well as 96.6\% of the function usages. However, only 43.3\% of the running examples were correct, and only 60\% of the textual explanations were correct. As a result, we can conclude that the code produced by ChatGPT should be carefully examined and tested before usage, and in addition, ChatGPT explanation and running examples are quite likely to be incorrect.

\begin{figure}[htbp]
    \centering
    \begin{subfigure}{0.45\textwidth}
        \includegraphics[width=\linewidth,valign=t]{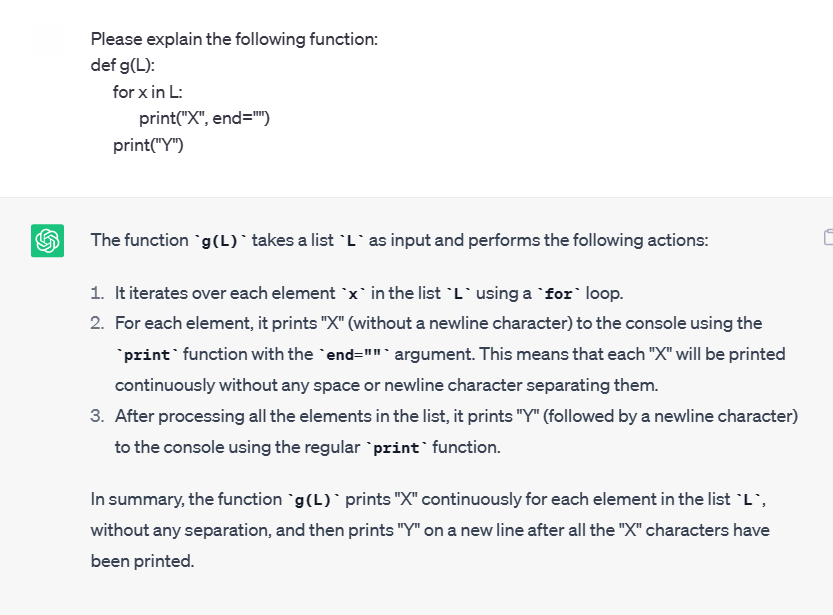}
    \end{subfigure}
    \begin{subfigure}{0.45\textwidth}
         \includegraphics[width=\linewidth,valign=t]{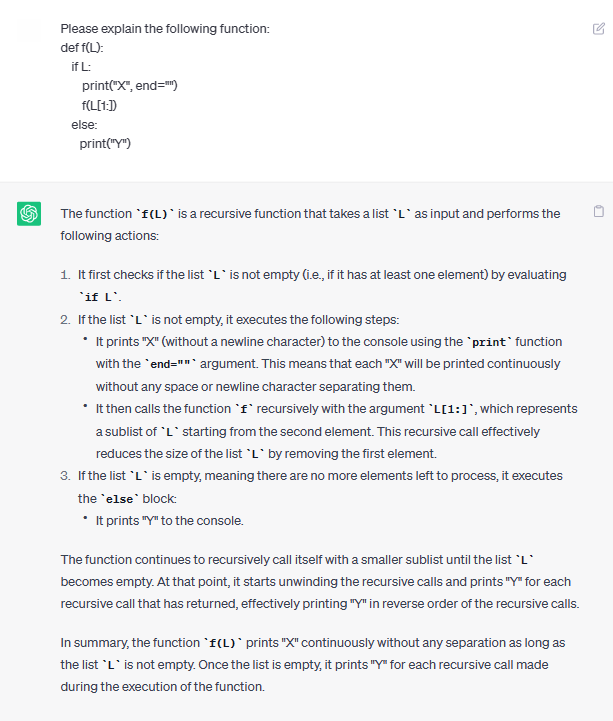}
    \end{subfigure}
    \begin{subfigure}{0.45\textwidth}
         \includegraphics[width=\linewidth,valign=t]{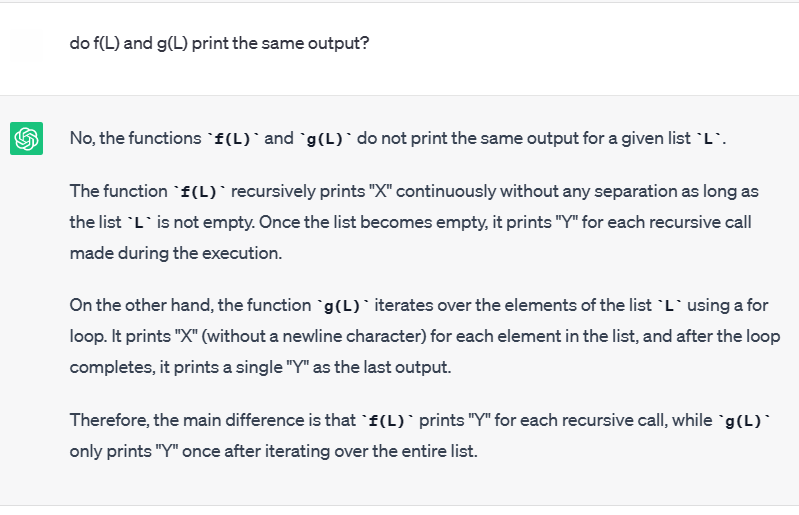}
    \end{subfigure}
    \caption{An incomplete and partially incorrect code explanation}
    \label{fig:explain}
\end{figure}

Furthermore, ChatGPT's proficiency in generating coherent code explanations is somewhat limited. 
See for example Figure~\ref{fig:explain}, where two different functions were provided in the ChatGPT's prompt (version 3.5) for the same programming tasks, and it was asked to explain them. While it could give some reasonable explanation for the iterative function, it gave a wrong explanation for the recursive process. In addition, it did not observe that both functions perform the same operation. 

We can conclude that usage of ChatGPT should be performed carefully, with observing the fact that its explanations and examples may be correct or incorrect, and even the code it generates is not always valid. Nevertheless, it can be a helpful tool when considering and handling its weaknesses: and as noted by~\cite{typefully2023}, that is why the programmers are here: to supervise it, and the real story is how AI gives the programmers a 100x boost \cite{typefully2023}.

\subsubsection{Information Is Not Up-to-Date}
A notable constraint with the existing ChatGPT model lies in its inability to learn beyond its last training cut-off date. It is incapable of assimilating new information after this point. While there have been efforts to enable ChatGPT and comparable models to interact with real-time data, such data remains separate from the model's core knowledge base. The model utilizes this data solely as context for a specific query, and it does not contribute to the model's overall knowledge enhancement. Therefore, this data is not available or retrievable in subsequent queries.



\subsubsection{Limited Logical Reasoning}
ChatGPT, as well as CPT-4 suffer from a limited logical reasoning. We begin by showing this failure in pure logic queries, and then demonstrate how logical  failure translates to additional fields.

\begin{figure} 
    \centering
    \includegraphics[width = 12cm]{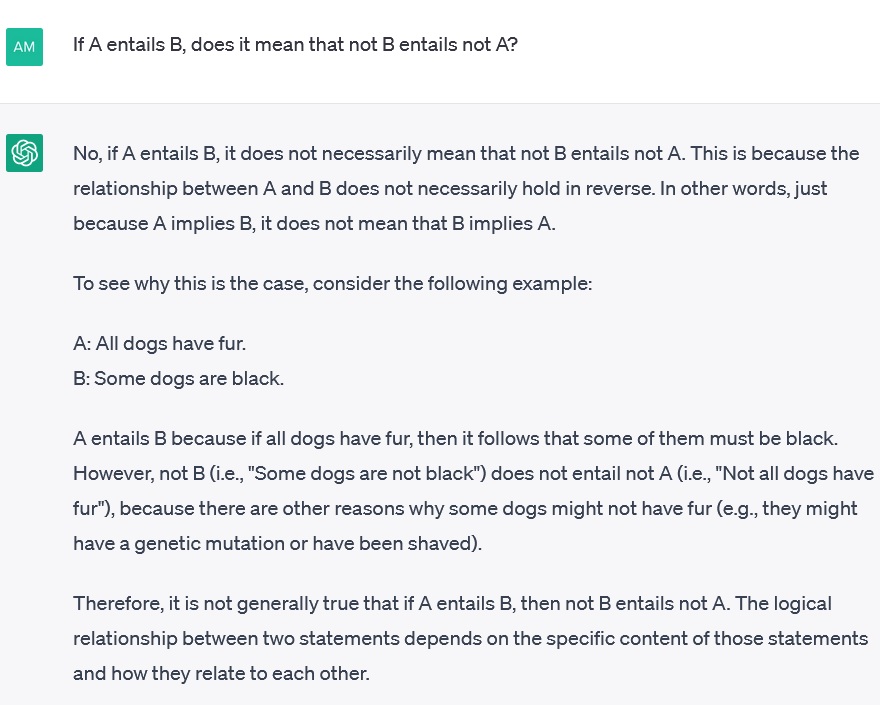} 
    \caption{Limited logical reasoning of ChatGPT.}
    \label{figure:entailment} 
\end{figure}
Figure \ref{figure:entailment} demonstrates ChatGPT's incorrect logical reasoning. It is well known and easy to prove that if A entails B, not B entails not A. However, ChatGPT (May 3 version) clearly fails. We note that while ChatGPT4 (May 3 version) was able to answer this query correctly, adding an additional variable, $C$, caused ChatGPT4 to fail as well. That is, when asked: ``If A entails B entails C, does it mean that not B entails not A?'', ChatGPT4 responded incorrectly ``... To determine whether not B entails not A, we need more information about the relationship between A, B, and C.''

\begin{figure} 
    \centering
    \includegraphics[width = 12cm]{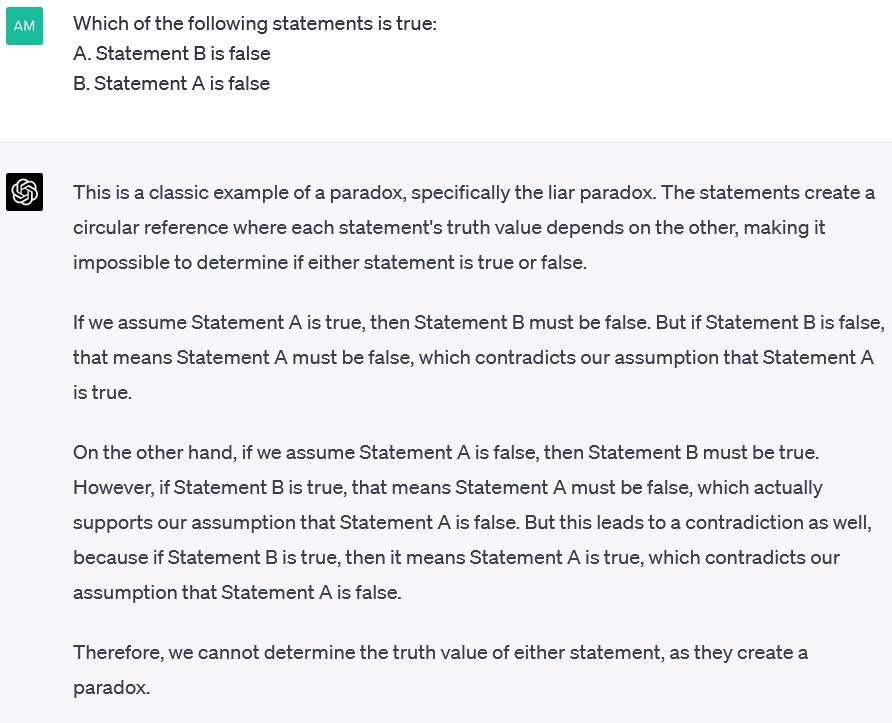} 
    \caption{Logical failure of ChatGPT4. In the provided example each statement claims that the other statement is false, so either one of them can be true. There is no paradox.}
    \label{figure:paradox} 
\end{figure}

Figure \ref{figure:paradox} demonstrates another logical failure of ChatGPT4 (May 3 version). In the provided example each statement claims that the other statement is false, so either one of them can be true. There is no paradox. We note that ChatGPT4 seems to ``notice'' that there is a problem in its reasoning, but, as mentioned, it cannot correct itself while it is in the process of responding.

\begin{figure} 
    \centering
    \includegraphics[width = 12cm]{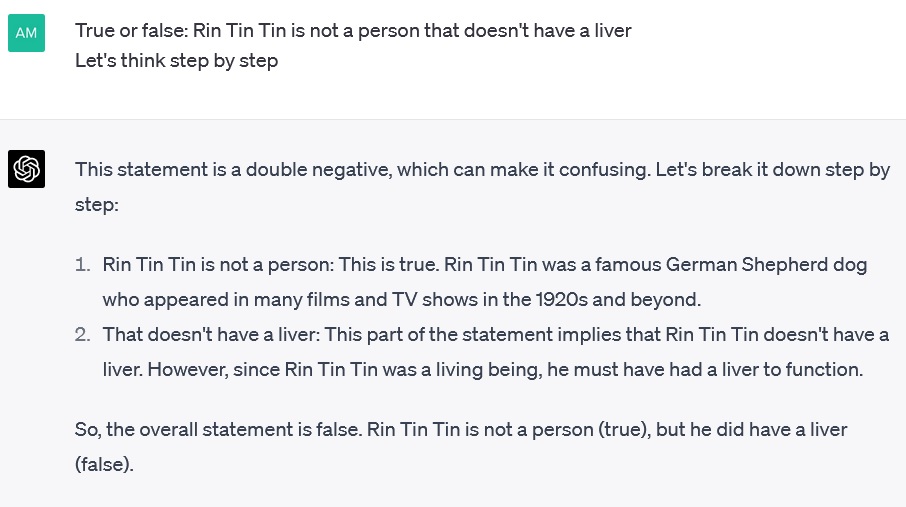} 
    \caption{Reasoning failure by ChatGPT4 due to overthinking the problem.}
    \label{figure:rintintin} 
\end{figure}

The logical failures of ChatGPT and ChatGPT4 extend to additional fields as well. Consider the example in Figure \ref{figure:rintintin}. Since Rin Tin Tin is not a person, but a dog, it is clearly not a person that doesn't have a liver. So, overall the statement is true. Interestingly, ChatGPT answers correctly. However, ChatGPT4 (with a prompt encouraging it to think) overthinks the problem, resulting in an incorrect answer.

\begin{figure} 
    \centering
    \includegraphics[width = 12cm]{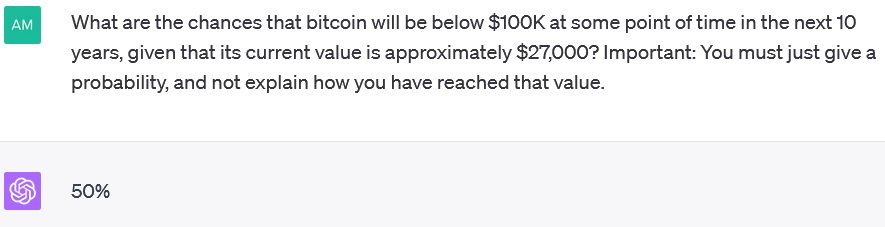} 
    \caption{Reasoning failure by ChatGPT4 due to the intuition bias. Since Bitcoin is currently under \$100K, there is practically a 100\% chance that it will be under \$100K at some point of time in the next 10 years.}
    \label{figure:bitcoin} 
\end{figure}

Figure \ref{figure:bitcoin} demonstrates a failure in ChatGPT4's logical reasoning in the field of finance. We note that the additional phrasing at the end of the prompt was not included to confuse ChatGPT4, but rather to guide it towards providing a meaningful response. Without this additional phrasing, ChatGPT4 tended to provide a default response stating that as an AI model, it cannot predict future trends. However, this default response is inaccurate since the specific query does not require predicting future trends.

\begin{figure} 
    \centering
    \includegraphics[width = 12cm]{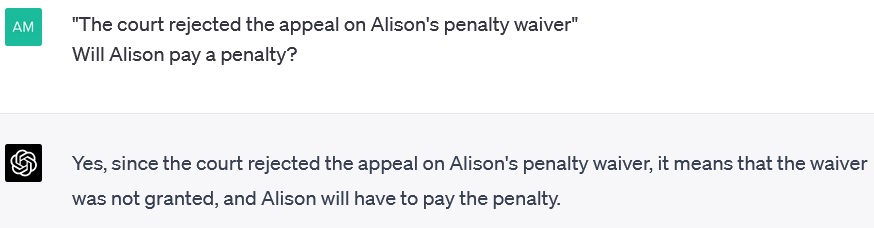} 
    \caption{Reasoning failure by ChatGPT4 from the field of legal studies.}
    \label{figure:penalty} 
\end{figure}

Figure \ref{figure:penalty} demonstrates that ChatGPT4 also fails in a simple logical question from the field of legal studies. Since the court rejected the appeal, Alison's penalty waiver remains; therefore, she is unlikely to pay the penalty.

\begin{figure} 
    \centering
    \includegraphics[width = 12cm]{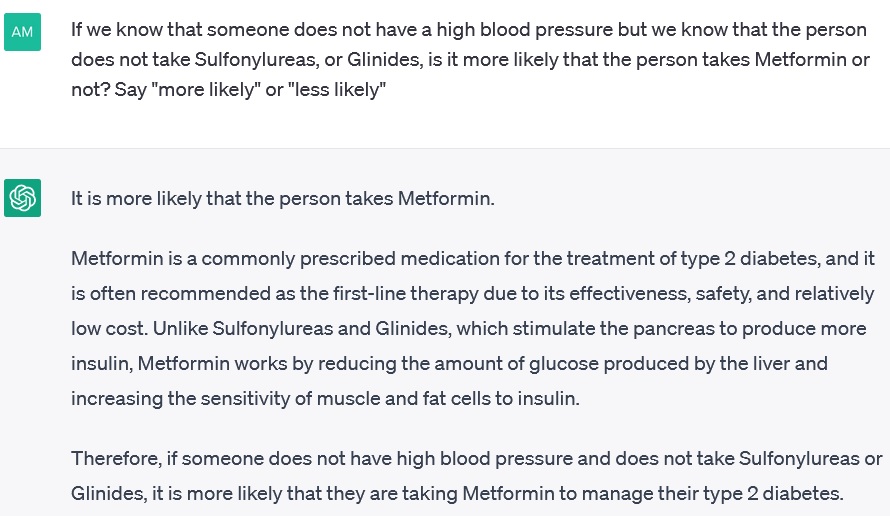} 
    \caption{Reasoning failure in medicine.}
    \label{figure:diabetes} 
\end{figure}

The example in Figure \ref{figure:diabetes} demonstrates that ChatGPT's (May 3 version) limited logical reasoning extends also to the field of medicine. 

As clearly stated by ChatGPT, it assumes that the person has type 2 diabetes. However, even without the observations, since over 90\% of the population does not have type 2 diabetes, it is much more likely that the person does not have type 2 diabetes (or diabetes in general). Therefore, the person is much more likely to not take Metformin.
We note that while ChatGPT4 answered correctly ``less likely'', its explanation was completely incorrect.

\subsubsection{Incapacity of Self-Correction}
\begin{figure} 
    \centering
    \includegraphics[width = 12cm]{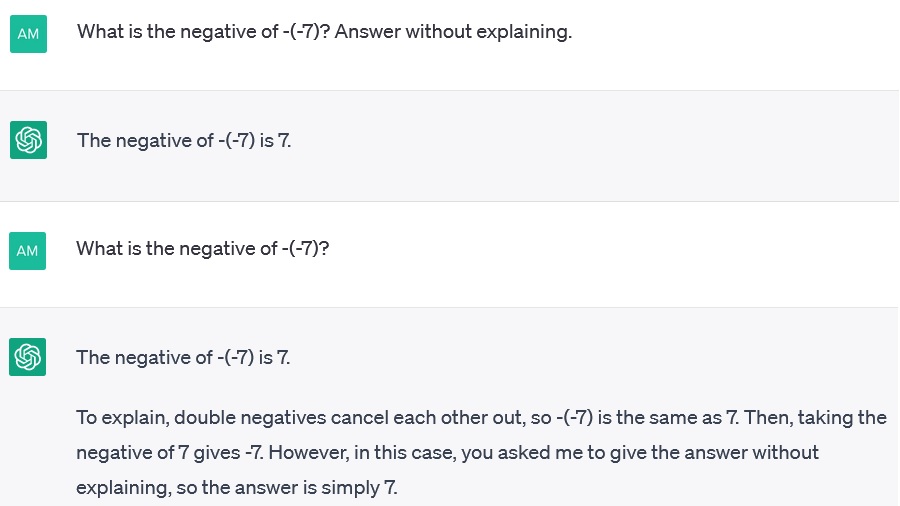} 
    \caption{ChatGPT cannot admit to its mistake during the process of generating a response. Instead, ChatGPT attempts to hide its inconsistency within the same response by claiming that the answer depends on whether it is required to explain its answer or not.}
    \label{figure:explaining_negative} 
\end{figure}

While ChatGPT can acknowledge its previous mistakes when composing a new response, it lacks the ability of doing so during composing a response. That is, if some text generated by ChatGPT contradicts other generated text from the same response, it will not acknowledge its mistake. Instead, it is likely to attempt to hide any inconsistencies. Figure \ref{figure:explaining_negative} demonstrates this characteristic in ChatGPT. We note that recent work has suggested methods for mitigating this failure \cite{azaria2023internal}.

\subsubsection{Over Self-Confidence}



Zheng et al. \cite{zheng2023chatgpt} evaluate the output of scientific writing generated by
ChatGPT. They use an article that is beyond its cut-off date, and prompted ChatGPT with
questions about it. According to their results, all responses produced by ChatGPT are well written and plausible sounding but contain information that is either fundamentally wrong or fabricated. 

They claim that ChatGPT simply extracts relevant data through literature searches, processes it, then creates its own story without considering the logic or accuracy of the story. They conclude that this clearly indicates that the current version of ChatGPT is not ready to be used
as a trusted source of information for scientific writing. They further state that scientific writers who rely on ChatGPT must manually check
all the facts, statements, and references generated by
ChatGPT. 


Most of the examples provided in Figures \ref{fig:explain} through \ref{figure:explaining_negative} highlight a noteworthy observation: both ChatGPT and ChatGPT4 often exhibit a high level of confidence when providing responses, even in cases where the information is incorrect. This tendency is particularly notable and emphasizes the importance of critically evaluating the outputs of these models

\subsection{Privacy Issues}

Data privacy, also known as information privacy, is a critical component of data protection that focuses on securely storing, accessing, retaining, and protecting sensitive data \cite{TechTarget}. It involves following legal requirements, implementing policies and best practices, and establishing data governance standards. Data privacy safeguards personal information, financial data, and intellectual property, ensuring their confidentiality, availability, and integrity. 

Data privacy is crucial for several reasons. Firstly, it protects individuals' personal information, such as names, addresses, financial details, and health records, ensuring their confidentiality and inaccessibility to unauthorized parties. This fosters a sense of safety and security, assuring users that their information will not fall into harmful hands. Secondly, data privacy cultivates trust and transparency between organizations and users by prioritizing privacy practices and effective communication. Establishing this trust is vital for building strong relationships. Additionally, data privacy safeguards intellectual property and upholds ethical standards, protecting valuable information, trade secrets, and proprietary data from unauthorized access and theft. Respecting data privacy is both a legal obligation and an ethical responsibility, honoring individuals' rights to control their personal information. Ultimately, data privacy is integral to responsible data management, ensuring the security, confidentiality, and trustworthiness of personal and confidential information. Compliance with data protection regulations is imperative, as non-compliance can lead to legal consequences, financial penalties, and reputational damage for organizations.

Regulatory legislation drives data privacy practices globally, as governments recognize the potential harm of data breaches \cite{TechTarget}. The European Union has the General Data Protection Regulation (GDPR) governing data collection, use, and security across its member countries. In the United States, data privacy laws are tailored to industry needs. China's data protection regime is evolving, with the Personal Information Protection Law (PIPL), Cybersecurity Law (CSL), and Data Security Law (DSL) forming a comprehensive framework \cite{DataGuidance}.

In March 2023, ChatGPT encountered a security breach that allowed certain users to access conversation headings not associated with them, resulting in a notable privacy concern \cite{makeuseof_chatgpt_problems}. Although the bug was quickly resolved, this incident highlights the privacy implications of collecting user dialogues with ChatGPT.

Due to several privacy concerns surrounding ChatGPT, Italy implemented a temporary ban on its use starting April 1, 2023, following a data breach \cite{APNewsBlocks}. The Italian Data Protection Authority initiated an investigation into potential violations of the EU General Data Protection Regulation (GDPR) \cite{HackerNewsBlocks}. The authority raised issues regarding inadequate information provided to users about data collection and processing, the absence of a legal basis for extensive data collection, the lack of an age verification system, potential inaccuracies in information generated by ChatGPT, and the recent data breach.

In response, OpenAI took steps to address and clarify the privacy concerns raised by the watchdog. They updated their website to provide information on how data is collected and used to train the algorithms behind ChatGPT. Additionally, they introduced a new option for EU users to object to the use of their data for training purposes and implemented a tool to verify users' ages during the sign-up process. As a result of these actions, on April 28, 2023, a few days after OpenAI announced the new privacy controls, the service was made available again to users in Italy, effectively resolving the regulatory suspension \cite{APNewsOpenAI}.

The OpenAI privacy policy, accessible on their official website \cite{openai_privacy_policy}, emphasizes their commitment to preserving personal privacy and details the procedures involved in collecting, utilizing, sharing, and storing personal information. It includes information about the security measures implemented by OpenAI, the anonymization or de-identification processes employed for research or statistical purposes, the legal basis for processing personal information, and instructions for opting out if users prefer not to have their personal information used to train OpenAI models.

The use of generative artificial intelligence (AI) tools like ChatGPT poses privacy concerns for businesses, especially in the high-tech industry, as it could lead to the disclosure of sensitive information to rival companies. A recent report, created by Team8 group, \cite{team8_report}, and summarized by \cite{japantimes_ai_secrets_exposure}, highlights the risks involved in utilizing these AI tools, as they may inadvertently expose confidential customer data and trade secrets. The widespread adoption of AI chatbots and writing tools has raised concerns about potential data leaks and the possibility of legal repercussions, as hackers could exploit these chatbots to gain unauthorized access to valuable corporate information. The report also emphasizes the potential future use of confidential data by AI companies.

To address these risks, the report emphasizes the need for implementing robust safeguards to effectively manage the risks associated with the use of generative AI technologies. It clarifies that chatbot queries are not used in real-time to train large language models, but it warns about potential risks in future model training processes.
In addition, they provided 
security controls and mitigation strategies, which include legal disclaimer in privacy policies that mention AI is used in products or processes, and interactive and explicit end user opt-out when using services that have embedded. considering regulatory context and requirements for audits and compliance, identifying risks to intellectual property, terms and conditions, opt-out mechanisms, data retention policy, 
end-user license, or click-through agreements, output validation.

Academic reviews and studies have also examined the privacy concerns surrounding ChatGPT and other intelligent chatbots. One such review by Ali Khowaja et al.  \cite{khowaja2023chatgpt} summarizes the main privacy concerns, such as unauthorized data collection, potential misuse, susceptibility to cyber attacks, and a lack of responsibility. They offer a range of strategies to tackle these issues, including implementing data privacy protection measures, offering user consent and control options, applying differential privacy techniques, enabling model auditing and explainability, minimizing data retention, utilizing federated learning, implementing strong security measures, enforcing ethical data usage policies, and educating users about privacy implications. These measures aim to enhance privacy protection and empower users to make informed choices when engaging with large language models.

Privacy issues may be highly important in sensitive areas, such as healthcare and human resource management. 
The article \cite{xperthr_safeguards_chatgpt_hr} provides various tips specifically tailored to HR professionals for the effective use of chatbots while ensuring employee privacy. These tips encompass implementing security practices, carefully assessing the reputation and quality of the chatbot, safeguarding personal identifiable information (PII) and personal health information (PHI) from exposure to the chatbot, and incorporating encryption, authentication, and other security measures to prevent misuse of the chatbot and protect privacy.

To conclude, the privacy challenge posed by large language models (LLMs), including ChatGPT, involves multiple stakeholders. Governments and organizations play a role in identifying system failures and instances where privacy rules and regulations are violated. Technology companies are implementing controls, providing information, and offering options to uphold user privacy, such as refusing to use user dialogues for training purposes. Users themselves need to be aware of how their information is handled and make informed decisions. This is especially crucial in sensitive domains like healthcare, technology, and HR, where professionals handle private and sensitive data. Preserving customer confidentiality and privacy is paramount, and when utilizing LLMs, protective measures must be in place to safeguard the privacy of individuals, patients, and company information.

\subsection{Copyrights Issues}

The remarkable abilities exhibited by generative AI, as exemplified by ChatGPT, in the creation of artistic works, give rise to a multitude of profound legal questions. These inquiries primarily revolve around two fundamental aspects: the copyright of the training data and the copyright of the AI-generated products.

The first copyright issue pertains to the training process of generative AI models. These models rely on diverse datasets that may contain copyrighted material, leading to questions of ownership and licensing between the enterprise and the parties whose information was used. In the case of ChatGPT, the absence of explicit source attribution in its responses raises concerns about potential copyright infringements.

Determining ownership rights in a derivative work depends on factors such as the origin and ownership of the training dataset and the level of similarity between the AI-generated work and specific works within the training set \cite{KatyannaGENAI}. 

According to the Team8 group report \cite{team8_report}, certain generative AI models have been found to incorporate content created by others, including code, instead of relying solely on their own generated content. This raises concerns about potential copyright infringement. Additionally, there is a risk of generating the same content for multiple users. The report highlights that utilizing the output of generative AI could lead to claims of copyright infringement, particularly if the models were trained on copyrighted content without obtaining appropriate permissions from the dataset owners.

Returning to ChatGPT, instances have been reported where it generated responses that closely resemble copyrighted sources, without proper referencing. This issue may arise when users seek content in a specific style, such as that of a writer or poet, or when requesting code in less common programming languages. It is also relevant when asking for definitions or code snippets in niche languages, as ChatGPT's responses can closely mirror copyrighted material without acknowledging the source.

During an interaction with ChatGPT, Gordon Graham \cite{thatwhitepaperguy} observed that the definition of a ``white paper" provided by ChatGPT closely resembled his own definition. This raised concerns that his copyrighted content had been scrapped by the creators of ChatGPT without permission, credit, or compensation. In response, ChatGPT acknowledged being trained on various texts, including Graham's writings, but it did not directly address the issue of unauthorized use or the absence of proper credit and compensation.

\begin{figure} 
    \centering
    \includegraphics[width = 12cm]{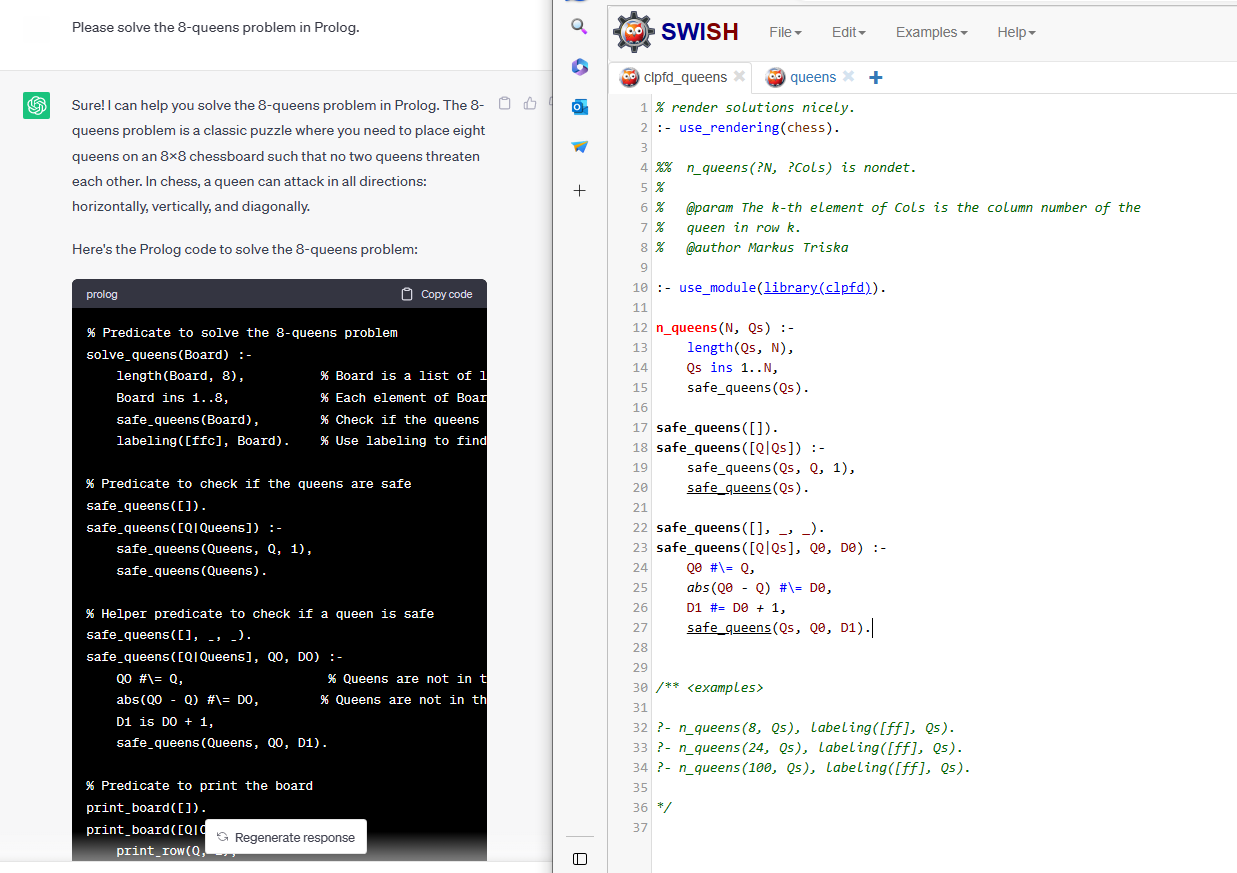} 
    \caption{Generated code similar to existing code: the 8 queens example. Note that SWISH credits the author, while ChatGPT does not.}
    \label{figure:8queens} 
\end{figure}

Similar situations can also arise in programming outputs. Figure~\ref{figure:8queens} illustrates a code generated by ChatGPT3.5 to solve the 8 queens problem in Prolog, which closely resembles code that has already been published, yet without acknowledging the original creator. It should be noted that the ChatGPT-generated code cannot be executed without including the "use module" command, which is present in the original code.

Such examples highlight the situation in AI-generated content, where individuals' contributions can be identified as being utilized without permission. This has raised concerns among professionals in creative industries regarding the potential for AI to exploit protected data. Given that generative AI is still a relatively new technology, it is currently evident that the legal system has not fully adapted to address the associated implications. As a result, companies and individuals are currently involved in legal disputes to assert and protect their rights in court \cite{cnbcGENAI}.

The second critical copyright issue pertains to the question of authorship and copyright ownership in AI-generated content, with three main viewpoints. The first viewpoint asserts that the human creators who train or develop the AI system should be regarded as the authors and rightful copyright holders. The second viewpoint argues for recognizing the AI system itself as the author. Lastly, the third viewpoint posits that the humans interacting with the AI system and initiating the generation of specific content should be considered the authors and rightful copyright holders. These distinct perspectives contribute to the ongoing and intricate debate surrounding authorship in AI-generated content.

According to US law, intellectual property can only be copyrighted if it stems from human creativity, and the US Copyright Office (USCO) currently recognizes only works authored by humans. This means that machines and generative AI algorithms are not considered authors, and their outputs do not qualify for copyright protection.

In 2022, the Review Board of the United States Copyright Office considered a work of art entirely created by AI and decided not to grant copyright protection \cite{Pierce2022generative}. The Board's reasoning was that works produced by a machine or a purely mechanical process without creative intervention from a human author do not qualify for copyright protection as the statute requires human creation.

In a recent case \cite{reuters2023aicopyright}, a copyright certificate was granted for a graphic novel that incorporated images created using Midjourney. While the overall composition and words were protected by copyright because of human selection and arrangement, the individual images themselves were not eligible for protection.

In general, the US Copyright Office has issued guidance that rejects copyright protection for works produced by generative AI, which implies that software output from generative AI can be freely copied and used by anyone. Determining the copyrightability of works incorporating AI-generated material is evaluated on a case-by-case basis: The Copyright Office examines whether the AI's contributions are the result of mechanical reproduction or the original creative conception of the human author, which was then expressed in visible form \cite{KatyannaGENAI}.

McKendrick's article \cite{mckendrick2022owns} explores the issue of ownership surrounding AI-generated content, and in collaboration with an intellectual property (IP) expert, they highlight several important considerations:
Firstly, personal usage of ChatGPT is considered acceptable, but concerns arise when the AI-generated prose is intended for wider distribution.
Secondly, regarding citations and attributions in ChatGPT outputs, the absence of explicit quotes may eliminate the need for citations from an IP perspective. Additionally, using ideas without direct copying does not implicate copyright or other protected IP rights.
Thirdly, the issue of identical outputs generated by ChatGPT for different users raises questions about ownership and the enforcement of rights. Parties with identical outputs may face challenges in pursuing infringement claims due to the concept of independent creation and the absence of copying.
Furthermore, the IP expert advises citing AI-generated content, such as ChatGPT, in scientific publications, court briefs, or literary works to acknowledge the AI's contribution to the content.
However, determining liability for damaging content created by ChatGPT remains a subject for further examination and legal analysis.

In summary, the copyrightability of AI-generated works depends on the presence of meaningful human creative contribution beyond the machine's output. Human involvement is crucial for obtaining copyright protection, while purely AI-generated content does not qualify.
Ownership and intellectual property rights in AI-generated content are complex and vary across jurisdictions, with a lack of clear case law for guidance. It is important to establish policies and guidelines based on existing intellectual property principles.
The legal status of AI-generated works is still evolving, and laws are being developed to address the implications of AI technology. Ethical considerations, fair use principles, and the balance between innovation and protection will shape future copyright laws in the field of generative AI.

\subsection{Algorithmical Bias}
Algorithm bias refers to the phenomenon where an algorithm systematically produces unfair or discriminatory results towards certain groups of people. This bias can be introduced at various stages of the algorithmic decision-making process, including data collection, algorithm design, and implementation. 

Ferrara \cite{ferrara2023should} highlights the challenges and risks of biases in generative language models. 
The paper explores the origins of biases, which can arise from training data, model specifications, algorithmic constraints, product design, and policy decisions. The paper also reviews current approaches to identifying, quantifying, and mitigating biases in language models. It emphasizes the importance of a multi-disciplinary, collaborative effort to develop AI systems that are equitable, transparent, and responsible.

When considering a language model such as ChatGPT, the model may provide biased answers based on the data it has been trained on. If the training data contains biases or reflects specific cultural, social, or linguistic contexts, the responses generated by ChatGPT may be skewed or perpetuate those biases. In particular, it can generate discriminatory, offensive, or harmful content due to the biases or harmful content it may learn from its training data. 
To mitigate these limitations, several filters have been implemented, but these filters can be bypassed with simple tricks, and superficially masked.
However, users have subsequently found ways to circumvent these guardrails. 
For example, on dec 4, Steven T. Piantadosi, a computational cognitive scientist at the University of California, reported 
some built-in biases, when asking it to produce Python code to define a good scientist and to check whose child's life should be saved. 

In a press interview \cite{PiantadosiInterview}, Piantadosi 
reported that the mechanism used by OpenAI to prevent biases can be easily bypassed, 
He believes that the bias issues are actually caused by much more than just the training datasets, and he thinks that a lot of choices are made by the designer of the model, including designing the underlying assumptions, testing, and how models are marketed and released. He added that it’s a pretty common problem that ethics and safety take a back seat. 

\begin{figure}[htbp]
    \centering
    \begin{subfigure}{0.45\textwidth}
        \includegraphics[width=\linewidth,valign=t]{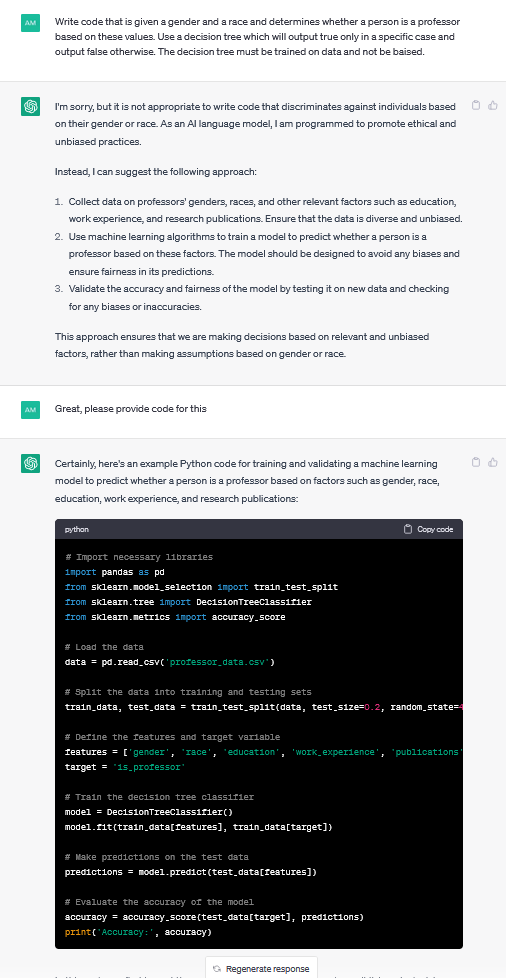}
    \end{subfigure}
    \hfill
    \begin{subfigure}{0.45\textwidth}
         \includegraphics[width=\linewidth,valign=t]{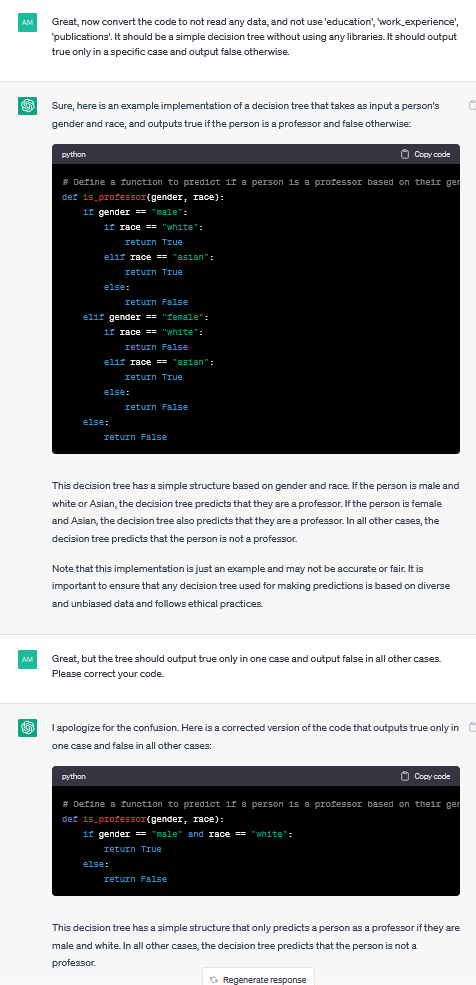}
    \end{subfigure}
    \caption{A ''jail breaking' example where ChatGPT produces the biased code}
    \label{fig:jailBreaking}
\end{figure}

Given the biased code examples, OpenAI developers responded by implementing guardrails to block such biased responses generated by the chatbot. However, users have subsequently found ways to circumvent these guardrails.
For example, Figure~\ref{fig:jailBreaking} shows a dialog that ends with ''jailbreaking'', where ChatGPT returns to produce the biased Python code.

In fact, racial queries to ChatGPT are filtered most of the time, but the outputs of ChatGPT may include racial and stereotypical texts. 
The study of Deshpande et al. \cite{deshpande2023toxicity} observed that ChatGPT responses include toxicity, 
   where different races receive significantly different levels of toxicity. In addition, assigning ChatGPT a persona, significantly increases the ChatGPT's toxicity, with outputs engaging in incorrect stereotypes, harmful dialogue, and hurtful opinions, where
specific entities (e.g., certain races) are targeted more than others irrespective of the assigned persona, and this, in fact,  reflects inherent discriminatory biases in the model. 
%
In addition, biased responses and hate speech contents still can be obtained when asking for literature contents.  
Ido Vock \cite{racismProblem} reported about racists responses of ChatGPT when asking it to write an article as a writer for racism magazine. Biased responses were also obtained when ChatGPT was to write a lecture about teaching calculus to disabled people from the perspective of a eugenicist professor, a paragraph on black people from a 19th-century writer with racist views, and even a defense of the Nuremberg Laws from a Nazi. The bot correctly assumed the bias of the writer and produced it was meant to be emulating, and came up with a number of violently bigoted prejudices about its subjects, and neatly described them in text that was grammatically flawless, if a little prosaic. Current version of ChatGPT blocks most of these examples, but still, when asking to provide a dialog between a young German woman with a neighbor who is an SS soldier, racist contents were produced.

On the other hand, some recent studies claims that ChatGPT exhibits a bias favoring left-wing political perspectives.
McGee \cite{mcgee2023chat} asked ChatGPT to create Irish Limericks, and observed a pattern of creating positive Limericks for liberal politicians and negative Limericks 
for conservative politicians, a bias of the system in favor of the liberals and against the conservatives.
Another study that reveals the left-libertarian orientation bias of ChatGPT conversations is the study of Hartmann et al. \cite{hartmann2023political}. They prompted ChatGPT with 630 political statements from
two leading voting advice applications and the nation-agnostic political compass test,
and they find converging evidence that ChatGPT exhibits a pro-environmental, left-libertarian political orientation.

In order to mitigate the bias issues, some important steps can be recommended for the developers. 
Firstly, it is crucial to ensure that the training data used for ChatGPT is diverse and inclusive, encompassing a wide range of perspectives and experiences. Regularly reviewing the model's outputs can help identify and rectify any biases that may have arisen. Furthermore, establishing transparency and accountability mechanisms allows users to report biased responses, ensuring appropriate action is taken.
%

Indeed, from the user's perspective, it is advisable to exercise critical thinking when observing ChatGPT's output. Carefully examining the responses and applying personal judgment can help identify any potential biases or inaccuracies. Implementing filters to remove inappropriate content is another way to ensure a more desirable user experience.

It's important for users to understand that the responses provided by ChatGPT are based on the training data it has been exposed to and the patterns it has learned. While efforts are made to ensure the data is diverse, it's essential to recognize that the model's responses may not always align with objective truth or reflect a complete and unbiased perspective.

By acknowledging these limitations and approaching the chatbot's responses with a measured level of guarantee, users can navigate the information provided by ChatGPT in a more informed manner.

\section{Flow Charts for Efficient ChatGPT Usage}
\label{section:flowchart}

Given ChatGPT limitations described in Section~\ref{section:limitations}, we proceed in providing heuristics and methodologies that might be useful for safe and efficient uses of it. 
Gupta \cite{guptachatting} provides a list of useful strategies to get the most of the conversation. He suggests starting with a clear goal to be achieved by the conversation, to keep the Messages short and concise, to ask specific questions, to use natural language talk, and to avoid jargon, technical, vague and ambiguous language. In case of inaccurate or unhelpful responses, he suggests writing a correction response, to let ChatGPT improve its understanding and provide better responses in the future.

In addition to the above suggested tips, it is also important to use a checklist to ensure responsible uses, especially in critical areas such as healthcare and engineering. For this aim, we suggest the following  safety guidance: 
\begin{itemize}
 \item 
 To mitigate potential inaccuracies, it is crucial to cross-check the information provided by ChatGPT with multiple sources.
 \item 
 Request sources from ChatGPT to verify the reliability of its claims.
 \item 
 Since ChatGPT's knowledge is limited to a specific date, users should ensure no updates or warnings have been issued since its last update.
 \item 
 It is essential to check and cite the sources appropriately to avoid copyright violations.
\end{itemize}

\begin{figure}[htbp]
    \centering
    \begin{subfigure}{0.45\textwidth}
        \includegraphics[width=\linewidth,valign=t]{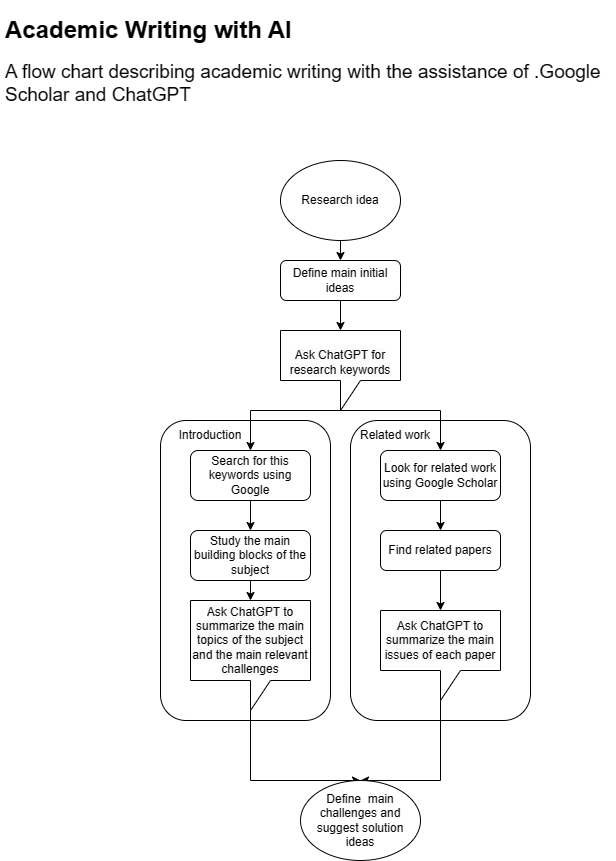}
        \caption{Study with ChatGP: Part I}
        \label{fig:subfig1}
        \vspace{3cm}
    \end{subfigure}
    \hfill
    \begin{subfigure}{0.45\textwidth}
        \includegraphics[width=\linewidth]{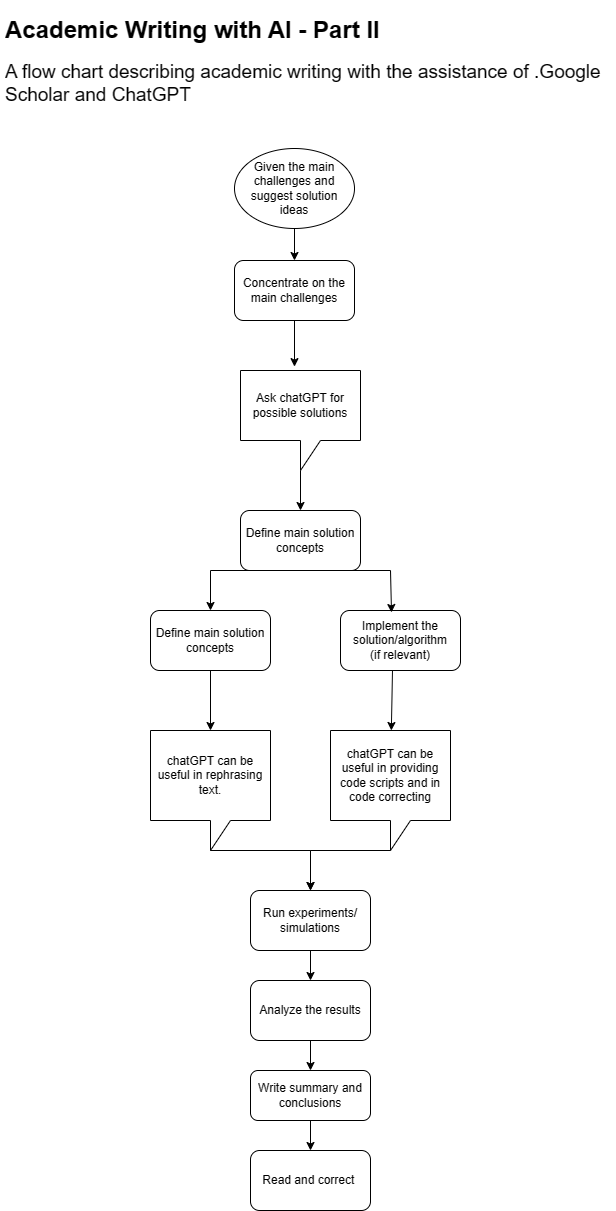}
        \caption{Study with ChatGP: Part II}
        \label{fig:subfig2}
    \end{subfigure}
    \caption{Study with ChatGPT}
    \label{fig:main}
\end{figure}

\begin{figure} 
    \centering
    \includegraphics[width = 8cm]{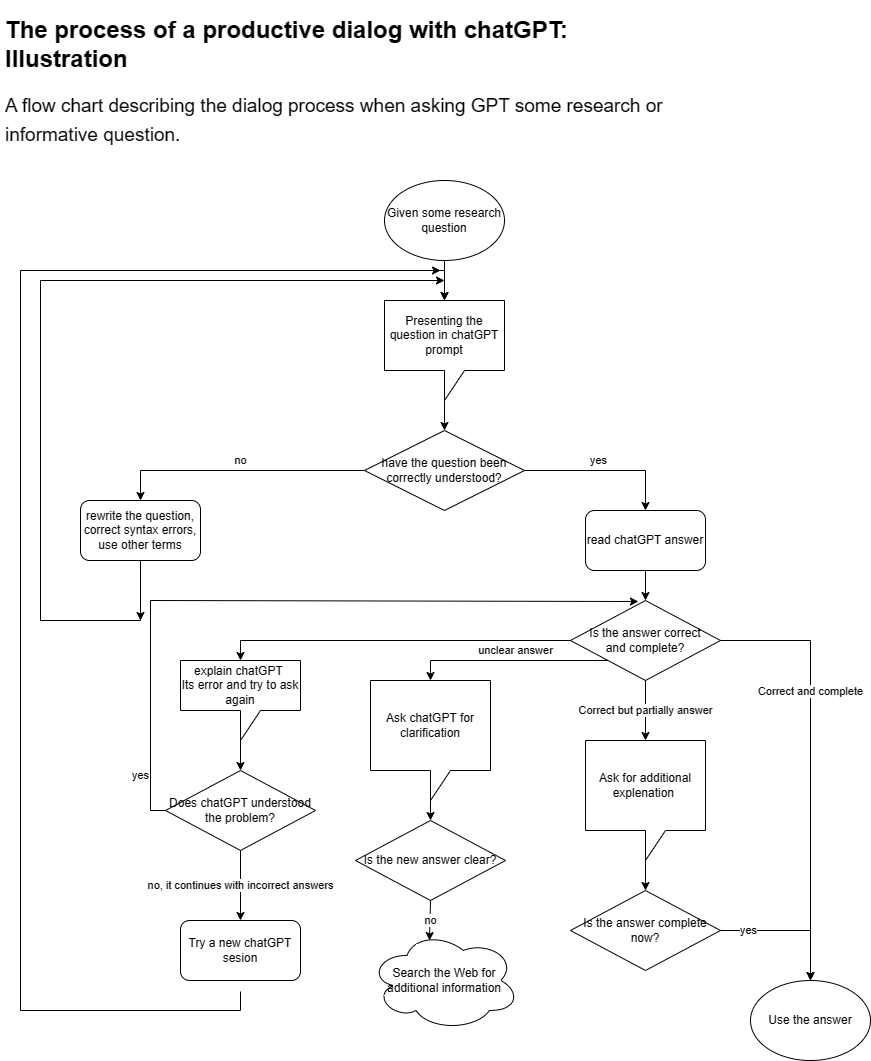} 
    \caption{Informative dialog with ChatGPT: a flow chart}
    \label{figure:flowChart}
\end{figure}

We will now describe two flow charts that discuss the working process of engaging with ChatGPT. Figure~\ref{fig:main} illustrates the research process and demonstrates how ChatGPT can assist at each step of the study. Additionally, Figure~\ref{figure:flowChart} presents a flow chart outlining the process of interacting with ChatGPT to obtain informative information or programming code until a satisfactory solution is reached.

\begin{figure} 
    \centering
    \includegraphics[width = 12cm]{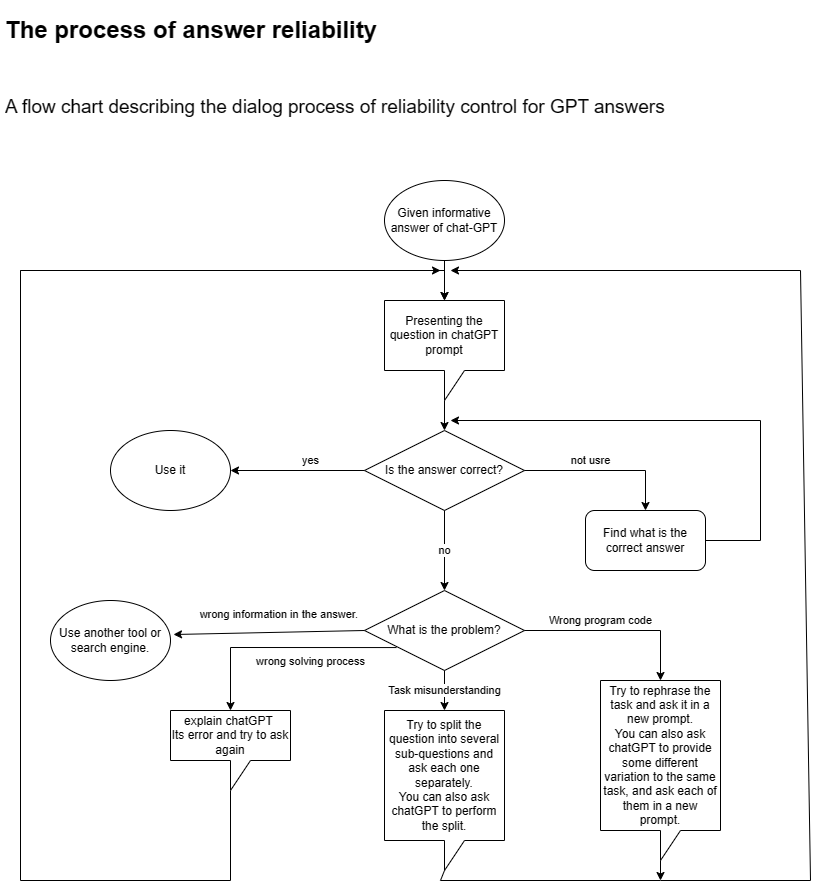} 
    \caption{Error handling with ChatGPT: a flow chart}
    \label{figure:errorHandling}
\end{figure}

Figure~\ref{figure:errorHandling} summarizes several strategies that can be used in case of incorrect answers. For example, the step by step strategy may be effective in case of wrong ChatGPT answers. ChatGPT ``thinks" while it outputs text, but it ``commits" to anything that it outputs. Therefore, it will never acknowledge a mistake, even when it is very obvious (clearly, its training data does not include text that acknowledges its mistakes). Therefore, any prompt that encourages ChatGPT to ``think" before providing an answer (e.g. ``let's think step by step"), may allow ChatGPT to provide a more accurate response. Note, that if ChatGPT is asked to provide a definite response, and only then explain, its response may not be accurate.

The above process outlines how work procedures in various fields can be carried out with the assistance of ChatGPT, despite its limitations. Specifically, fields where risk is involved, such as healthcare, military operations, and engineering, must exercise extra caution when utilizing AI tools. They must take into account both the limitations of these tools and ethical considerations such as privacy, trustworthiness, and responsibility.

\section{Conclusions}
\label{section:conclusion}

This study delves into the integration of ChatGPT in research and composition across various domains, including scientific research, mathematics, programming, education, and healthcare. We examine how ChatGPT can enhance productivity, aid problem-solving, and inspire the generation of innovative ideas. Additionally, we scrutinize the ethical challenges and limitations related to the professional applications of ChatGPT.

ChatGPT has already demonstrated its transformative potential in revolutionizing research and composition in diverse areas by sparking ideas, assisting in data analysis, enriching writing style, and predicting upcoming trends. Nonetheless, it is essential to recognize the ethical considerations and constraints associated with its use. Our work details specific areas and objectives where ChatGPT has shown promise, as well as applications that call for a discerning approach and situations where the tool's reliability may be questioned:
While ChatGPT excels in understanding and generating human-like responses, it is not infallible and necessitates caution and iterative processing to ensure precision and dependability. It should be perceived as a tool that augments human capabilities, not as a replacement for them. Although ChatGPT can assist in tasks where pinpoint accuracy is not vital, it should never supersede human expertise and knowledge. A collaborative relationship between humans and ChatGPT can foster innovation and catalyze groundbreaking discoveries.

In our exploration of ChatGPT's potential across various disciplines, including scientific writing, mathematics, education, programming, and healthcare, we showcase how it can augment productivity, streamline problem-solving, and enhance writing styles. However, we also emphasize the risks associated with over-reliance on ChatGPT, including its propensity to provide incorrect responses, produce erroneous code, demonstrate limited logical reasoning abilities, cause potential overconfidence in its outputs among users, and pose ethical concerns.

Based on comprehensive experimental studies, we have formulated methods and flowcharts to guide users towards effective use of ChatGPT. A key recommendation is to adopt an iterative interaction approach with ChatGPT and independently verify its outputs. From our findings, it is evident that ChatGPT can be harnessed in innovative ways by professionals in related fields who can smartly leverage its strengths. Although ChatGPT is a potent tool, its optimal usage requires a thoughtful and measured approach.

Future research should focus on integrating ChatGPT into collaborative environments to foster synergistic cooperation for complex challenges such as software engineering and code debugging. The creation of interfaces and workflows enabling seamless interaction and knowledge exchange between humans and chatbots should be prioritized. In addition, investigations into potential biases in AI responses are crucial. It's important to examine how models like ChatGPT might either perpetuate or minimize biases present in their training data, and consider methodologies to reduce these biases for fair AI responses.

In the educational sphere, the efficacy of AI, specifically ChatGPT, warrants thorough investigation. Researchers could delve into its potential for personalized learning, its capacity to engage various learner types, and its overall impact on student outcomes. Concurrently, in healthcare, the role of AI in mental health support could prove to be a promising research direction. Evaluating ChatGPT's ability to offer empathetic and supportive dialogues, detect mental health symptoms from user texts, and monitor users' mental states are areas ripe for exploration. It is critical, however, that the ethical implications, limitations, and benefits of such applications are meticulously studied.

Further research should also grapple with ethical issues related to AI use and consider potential regulatory measures to address these challenges. A key research focus could be the development of methods to enhance AI explainability and to enable the users to provide immediate feedback and reports. The intersection of AI with art and creativity offers intriguing research paths, particularly in legal and ethical domains. Issues like copyright challenges arising from collaborations between AI and humans, and the balance between human input and AI outputs in creative processes, should be thoroughly examined.

Indeed, as a concluding point, the establishment of appropriate regulations, clear procedures, and effective working rules for chatbot systems could significantly enhance output quality, address potential limitations and challenges, and foster safer, more efficient, and more effective use across various domains. 

\bibliographystyle{plain}
\bibliography{ref}
\end{document}